\documentclass[pre,twocolumn,reprint,amsmath,amssymb,superscriptaddress,showpace,superscriptaddress,groupedaddress,english]
{revtex4-2}

\usepackage{graphicx}
\usepackage{subfigure}

\usepackage{mathtools}

\usepackage{lineno}
\usepackage{times}
\usepackage{graphicx}
\usepackage{epstopdf}
\usepackage{subfigure}
\usepackage{sidecap}
\usepackage{amsmath}
\usepackage{mathtools}
\usepackage{appendix}

\newcommand{\etal}{{et al}. }
\newcommand{\ie}{{i}.{e}., }
\newcommand{\eg}{{e}.{g}. }
\newcommand{\veC}[1]{\boldsymbol{\vec{#1}}}

\begin{document}
\title{Slip flow regimes in nanofluidics: a universal superexponential model}

\author
{Mohammad Aminpour,$^{1,2}$, Sergio Andres Galindo Torres,$^{3,4\ast}$ Alexander Scheuermann,$^{2}$ and Ling Li$^{3,4}$}
\affiliation{
	$^{1}$Civil and Infrastructure Engineering Discipline, School of Engineering, RMIT University, Victoria 3001, Australia,\\$^{2}$School of Civil Engineering, The University of Queensland, Brisbane QLD 4072, Australia\\$^{3}$School of Engineering, Westlake University, Hangzhou 310024, China\\$^{4}$Key Laboratory of Coastal Environment and Resources of Zhejiang Province (KLaCER), School of Engineering, Westlake University, Hangzhou, China}

	\email{E-mail:  s.torres@westlake.edu.cn}

\begin{abstract}
	Many experiments have shown large flow enhancement ratios (up to $10^5$) in carbon nanotubes (CNT) with diameters larger than 5nm.  However, molecular dynamics simulations have never replicated these results maintaining a three-order-of-magnitude gap with measurements. Our study provides a generic model of nanofluidics for continuum slip flow (diameter$>\!\!3$nm) that fills this significant gap and sheds light on its origin. Compared to 140 literature cases, the model explains the entire range of experimental flow enhancements by changes of nanotube diameters and finite variations of interfacial energies. Despite large variations of flow enhancement ratios spanning 5 orders of magnitude in experimental results, the ratio between these data and corresponding model predictions approaches unity for the majority of experiments. The role of viscous entrance effects is discussed. The model provides insight into puzzling observations such as differences of CNTs and boron nitride nanotubes, the slip on low-contact-angle surfaces and massive functionalization effects. This study could advance our understanding of nano-scale transport mechanisms and aid the design of tailored nanomembranes. 
	
	
\end{abstract}
\maketitle

\section{Introduction}
Incredibly fast flows observed in carbon nanotubes (CNTs) \cite{majumder2005nanoscale,holt2006fast,secchi2016massive,huang2013ultrafast,whitby2008enhanced} have attracted much interest with potential applications in desalination, filtration and energy conversion \cite{das2014carbon,lee2011performance,srivastava2004carbon,siria2013giant}. High flow rates attributed to the slip condition on interfaces lead to significant flow enhancement ratios, $\varepsilon$ (the ratio between observed and classically predicted flow rates). 

However, the measured flow enhancement ratios in nanotubes varied substantially ($\varepsilon=1-10^6$). The controversy emerged from the early report of Majumder, \etal \cite{majumder2005nanoscale} who experimented on nanotubes of larger diameters ($d$=7nm) than those of Holt, \etal \cite{holt2006fast} ($d$=1.6nm) and yet measured greater flow enhancements than the latter by one to two orders of magnitude. Molecular Dynamics (MD) simulations only supported a portion of experimental results, mainly in a sub-continuum flow regime, typically for $d<1.66$nm \cite{thomas2009water, hummer2001water}. The simulations failed to explain the observed very intensive flow enhancements with $\varepsilon=10^3$-$10^5$ in larger nanotubes ($d>5nm$), although they were independently reported in several experiments \cite{majumder2005nanoscale, majumder2011anomalous,baek2014high,secchi2016massive,whitby2008enhanced,majumder2011mass,du2011membranes, lee2015carbon,zhang2015preparation}. 
The disagreement (a three-order-of-magnitude underprediction of $\varepsilon$ by MD simulations) led to scepticism on the accuracy of the measurements \cite{thomas2008reassessing, falk2010molecular,thomas2010pressure}. On the other hand, unambiguous measurements of water flow in individual CNTs provided further evidence of significant flow enhancements in relatively large tubes ($d=30-100nm$) \cite{secchi2016massive}.  With the measured flow rates far exceeding the numerical predictions, Secchi \etal posed more challenges on the MD simulations \cite{secchi2016massive}. Moreover, observing no flow enhancement for boron nitride nanotubes (BNNTs), which are crystallographically similar to CNTs, led to further complication concerning the slip flow behaviour \cite{secchi2016massive}. The insufficiency of MD models to identify the differences between CNTs and BNNTs \cite{suk2008fast,hilder2009salt} or to estimate the scales of such differences due to corrugation effects \cite{tocci2014friction,joseph2008carbon}, motivated the authors of Ref.  \cite{secchi2016massive} to link the phenomenon to physio-chemical factors, in a perspective beyond hydrodynamic theories. Overall, a three-order-of-magnitude mismatch has remained unresolved between the MD predictions and experimental results of nanotube flow enhancements \cite{kannam2017modeling}.

The range of MD (Lennard-Jones) intermolecular potentials are truncated at $1$ nm. The mechanisms beyond van der Waals (vdW) interactions including longer-ranged water molecule orientation, correlations in the hydrogen bonding network or proton hopping are poorly understood and too complex to be thoroughly simulated via MD models. However, supported by substantial experimental evidence, the \textit{pure} hydrophobic forces are much stronger and longer-ranged than the Lifshitz vdW force  \cite{israelachvili1982hydrophobic,meyer2006recent,donaldson2014developing}. Furthermore, small perturbations in carbon-water interactions could have drastic impacts on the nanotube transmissivity \cite{hummer2001water}, but no comprehensive study is yet performed.

A single-file or ring-style sub-continuum molecular mechanism for narrowest nanotubes ($d<1.66$nm) \cite{thomas2009water, hummer2001water}, and also the curvature effects are hypothesized as the origin of nanotube ultra-conductivity \cite{falk2010molecular}; however, none has explained high flow enhancements observed in larger nanotubes ($d>5$nm) \cite{majumder2005nanoscale, majumder2011anomalous,baek2014high,secchi2016massive,whitby2008enhanced,majumder2011mass,du2011membranes, lee2015carbon,zhang2015preparation}.

A few theoretical models have been proposed in search of a generalized slip flow model in nano-scales  \cite{wu2017wettability,shaat2019fluidity}. These models are based on limited MD data which are correlated linearly for estimating critical parameters, \eg slip lengths and viscosity. The contact angle, $\theta$ is normally used to quantify interfacial effects, however, it does not fully reflect the surface hydrophobicity \cite{israelachvili2015intermolecular} and is insufficient in reproducing MD interaction parameters \cite{leroy2015parametrizing}. Furthermore, these models link high flow enhancements to large $\theta$ ($\!>\!150^{\circ}$) which is incompatible with moderately hydrophobic nature of CNTs (with reported $\theta_{graphene}\!\approx\!70^{\circ}\!-\!90^{\circ}$ \cite{shih2012breakdown, van2017direct,rafiee2012wetting,kozbial2014study}, and $\theta_{CNT}\!\approx\!70^{\circ}$ \cite{ma2010dispersion}). Modifying classical hydrodynamics, another theory has corrected enhancement ratios from an order of $10^3$ to unity with some experimental data examined \cite{heiranian2019nanofluidic}.

Indeed, experiments \cite{majumder2005nanoscale, majumder2011anomalous,baek2014high,secchi2016massive} and MD simulations \cite{thomas2008reassessing, falk2010molecular,thomas2010pressure} have been challenging each other over the last fifteen years. A clear-cut model/theory for nanofluidic slip flow is yet to be achieved. Here using a continuum approach -proven to be valid for nanofluidics \cite{sparreboom2009principles,thomas2009water,bocquet2010nanofluidics}- incorporated with interface intermolecular potentials, we present a universal model of nano-scale continuum fluid transport for nanotubes and nanochannels. The model applies to the continuum regime where channel width $d\!>\!3nm$ and channel lengths are practically long in which entrance effects are not dominant.  In our model, a single dimensionless number uniquely characterizes the flow behaviour, \ie flow rate enhancement ratios (slip lengths). The model successfully explains the entire range of continuum flow experimental data including those unpredictable by previous MD simulations. This model explains the large experimental data scatters by changes in nanotube diameters where small perturbations or chemical variations of interface properties are also discussed. Our study also sheds light on the origin of a significant mismatch between MD simulations and experiments by incorporating the actual interfacial energy levels. We also discuss the significance of viscous entrance effects. 

Our method being already examined on complex and slip flow regimes \cite{aminpour2018pore,aminpour2018slip}, bypasses the limitations of computationally expensive MD models (in simulating larger dimensions, parametric sensitivity analyses, accurate force fields and complicated interplay of molecular features). Hence, the approach allows us to provide a more complete picture of flow behaviour in response to parametric variations, \ie interfacial effects (energies) and the system’s characteristic length scales. We also investigate the influence of different interfacial force functions and variations of their effective lengths to test the generality of the proposed slip flow model (see the Appendix \ref{App:A} for the details). 

\section{Modelling interface intermolecular interactions\label{sec:approach}}
 To consider the averaged intermolecular mechanism involved in nano-scale fluid slippage, we develop a numerical approach incorporated with well-established interaction potentials for interfaces based on nanoscale experiments \cite{israelachvili1982hydrophobic,donaldson2014developing}. The approach is potentially capable of including additional longer-range interactions not included in the combination of electrostatic double layer and vdW theories \cite{israelachvili2015intermolecular,donaldson2014developing}. A general interaction potential that best describes the hydrophobic interactions is developed as an exponential function of the effective interfacial energy $\gamma_i^{e\!f\!f}$  (positive for hydrophobic interfaces and negative for hydrophilic -hydrated- ones) \cite{donaldson2014developing,van2006interfacial,van1988interfacial} with a decay length of $\approx\!0.3\!\!-\!\!2nm$ \cite{donaldson2014developing}. For fully hydrophobic surfaces, $\gamma_i^{e\!f\!f}$ is equivalent to $\gamma_{i}$. For graphene as the substance for making CNTs, the (effective) interfacial energy in water is directly measured to be $\gamma_{i_(graphene)}\!\!=\!\!83\pm7$ $mJ/m^2$ \cite{van2017direct} with also similar values reported experimentally ($90$ $mJ/m^2$) \cite{wang2009wettability} and numerically ($94\!-\!99$ $mJ/m^2$) \cite{dreher2019calculation} (see Table \ref{tab:gammai} for a review).

Unlike the hydrophobic interactions, the hydrophilic ones are not well defined in term of interfacial energy. Negative interfacial energy means a tendency of the interface to expand indefinitely, \ie miscible media resulting in eventually dissolved interfaces with $\gamma_i$ approaching $0$, thus thermodynamically unstable. However, the interface may be kept in place by attractive physical interactions or chemical bonds to substrates (crystallization or strong covalent crosslinking) \cite{van2006interfacial,israelachvili2015intermolecular,donaldson2014developing}. Negative $\gamma_i^{e\!f\!f}$ accounts for solid-fluid attractive effects in hydrophilic interfaces with effective hydration forces resulting from the water ability to hydrogen bond and hydrate the surface. For a hydrophilic surface, observed effective interfacial energies are in the range of $\gamma_i^{e\!f\!f}=-0.5$ to $-15$ $mJ/m^2$ (\eg $-2.5$ for silica surfaces in water, and $-0.5$ to $-5$ $mJ/m^2$ for mica in different cation salt solutions), with the theoretical limit of the surface tension of water ($\gamma_l=72.8$ $mJ/m^2$ for water at $20^\circ C$) \cite{donaldson2014developing}.

\begin{figure*}[!htbp]
	\centering
	\includegraphics[width=\linewidth]{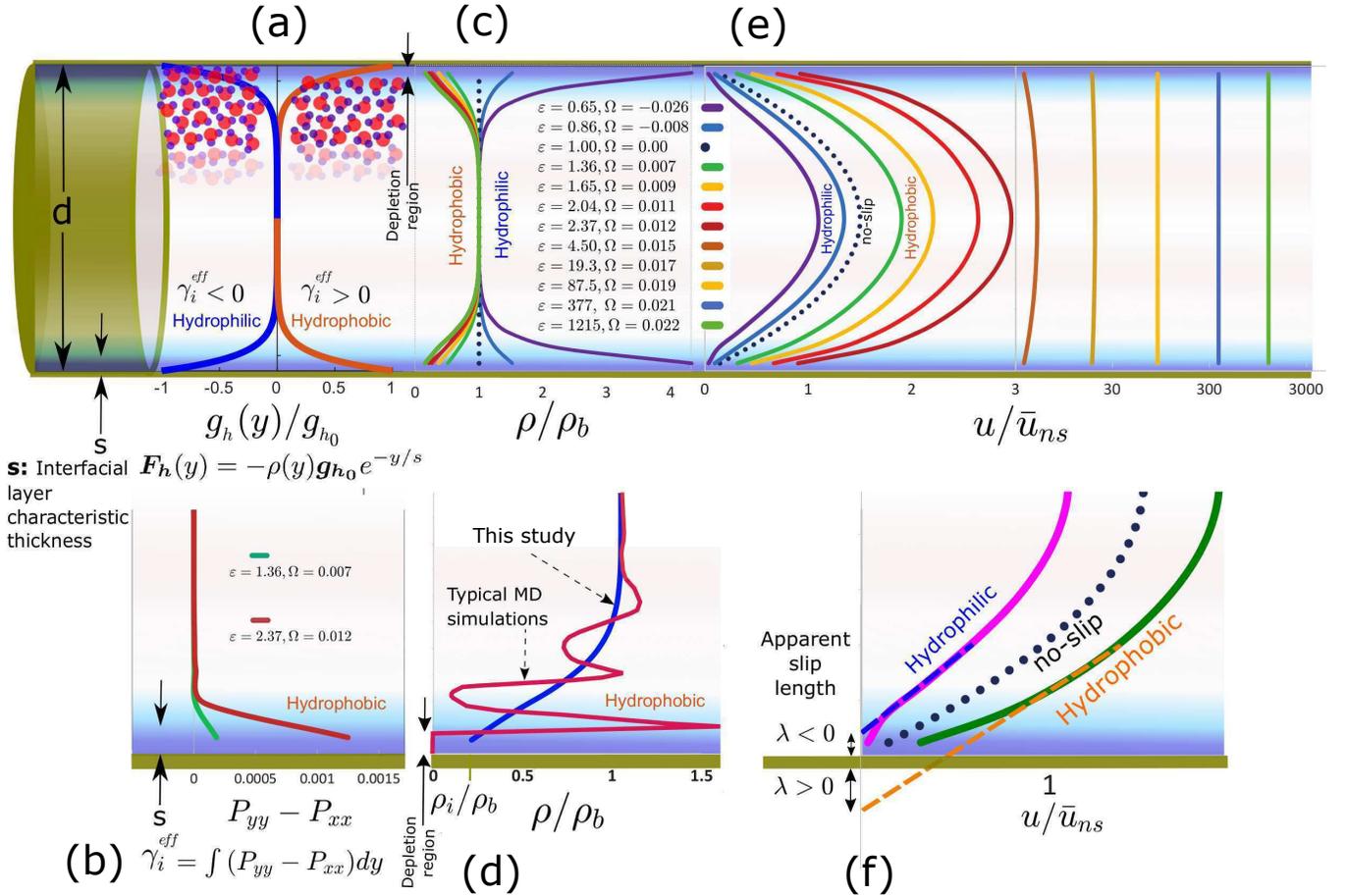}
	\caption {\textbf{Modelling slip flow characteristics in nanotubes.} 
		a) Intermolecular effects (combined vdW forces and effects beyond) are modelled using an interaction potential with an exponentially decaying, positive (hydrophobic) or negative (hydrophilic) force, corresponding to positive or negative $\gamma_i^{^{e\!f\!f}}$, respectively. The decay length of the force, $s$ ($\approx0.5nm$) is the characteristic thickness of the interfacial fluid layer. b) $\gamma_i^{^{e\!f\!f}}$ is calculated as the integrated difference of pressure components. c) The density variations (normalized by the bulk density, $\rho_b$) correspond to the dimensionless \textit{slip flow number}, $\Omega$ and the flow enhancement ratio, $\varepsilon$. d) Capable of incorporating the molecular effects in ranges longer than MD models, our approach retrieves the interface hydrodynamic features \eg slip velocities while upscaling the depletion
		and adsorption molecular layering.
		e) Enhanced and attenuated velocity profiles correspond to (f) positive and negative apparent slip lengths, respectively. The velocity jump takes place at the interfacial layer as also demonstrated by MD simulations \cite{ritos2014flow,joseph2008carbon}, leading to plug-like velocity profiles at higher flow enhancement ratios.
	}
	\label{fig:model}
\end{figure*}

Here, we develop an exponentially-decaying (hydrophobic or hydrophilic) solid-fluid intermolecular force function, $\boldsymbol{F_h}(y)=-\rho(y)\boldsymbol{{g}_{h}}=-\rho(y)\boldsymbol{{g}_{h_0}}e^{-y/s}$, where $\rho(y)$ is the fluid density, $y$ is the normal distance from the wall, $\boldsymbol{{g}_{h_0}}$ is the force strength factor (in units of acceleration) and $s$ is the decay length (see Fig. \ref{fig:model}-a). Such forces generate interfacial energies $\gamma_i^{e\!f\!f}$ (see Fig. \ref{fig:model}-b) calculated from simulations as the integration of pressure differences (see Appendix \ref{App:A} for methods). Upscaling the molecular-scale properties \eg depletion and adsorption molecular layering, this model facilitates the simulation of fundamental vdW forces and the longer-ranged effects. While the origin and exact molecular mechanisms of these forces remain unknown \cite{israelachvili2015intermolecular}, we retrieve their resultant effects -in scales of interfaces- and match it with accurately measured data in terms of $\gamma_i^{e\!f\!f}$ in nanoscopic experiments. We also show that the conclusions are independent of the solid-fluid force functions (exponential or constant) and the ranges of forces (short or long range), but mainly dependent on incorporating accurate interfacial energy levels (see Appendix \ref{App:A}, section 7).

Here, the definition of hydrophobic or hydrophilic surfaces is based on positive or negative $\gamma_{i}$, which is different from a sole $\theta$ criterion. $\gamma_{i}$ is directly related to the solid surface energy, $\gamma_s$ and $\theta$ via the Young's equation \cite{huhtamaki2018surface}:
\begin{equation}
\gamma_i=\gamma_s-\gamma_l \cos \theta\label{eq:young}
\end{equation}
where $\gamma_l$ is the fluid surface tension.

Our model is developed based on continuum simulations with surface-water interfacial interactions modelled using up-layer averaged molecular potentials. While averaging fluid layering at interface, the model averages molecular interactions including corrugations with their effects as variations in interfacial fluid density and viscosity. Also as shown in Ref. \cite{ho2011liquid}, both corrugation and fluid-solid interactions are affecting the slip where the responsible molecular signature appears to be the distribution of water molecules within the contact layer, coupled with the strength of water-surface interactions. This work also showed that the contact angle can change by variations of either lattice parameter or electrostatic parameter. Therefore, a change in corrugation can induce a variation in the macroscopic contact angle, despite the same intermolecular interaction parameters. One may then suggest that the average wetting property may inherently include the effects of corrugation when an up-layer averaging is maintained in simulations. Thus we suggest that the corrugation effects can be assumed as being implicitly included in our model.

\section{Flow rate variations and mechanisms\label{sec:results}}

The model predicts flow attenuation and enhancements (for hydrophilic and hydrophobic interfaces, respectively), indicated as deviations of the velocity profiles from the results under the no-slip assumption (Fig. \ref{fig:model}-e). Deviated velocity profiles extrapolate into (apparent) negative and positive slip lengths ($\lambda$), for hydrophilic and hydrophobic walls, respectively (Fig. \ref{fig:model}-f). With larger slip lengths, fluid velocities tend to plug-like profiles (Fig. \ref{fig:model}-e), as also demonstrated in MD simulations \cite{joseph2008carbon,ritos2014flow}.

The mechanism of such flow slippage (or inhibition) is attributed to fluid rarefaction (or densification) at interfacial layers next to hydrophobic (or hydrophilic) surfaces (Fig. \ref{fig:model}-c) extending to long-ranged molecule arrangements and hydrogen bonding incorporated with low (or high) fluid viscosity in the interface region. Such interfacial properties are supported by MD investigations demonstrating water depletion (and adsorption) next to hydrophobic (and less hydrophobic, \ie low-contact-angle) surfaces \cite{janecek2007interfacial,sendner2009interfacial}.  Furthermore, the thermodynamically driven low-density depletion regions on hydrophobic interfaces are shown by x-ray reflectivity measurements with the thickness in the order of water molecules ($2-4\AA$), disappearing with decreased hydrophobicity \cite{poynor2006water,mezger2006high}. Similar to our results, velocity jumps in the depletion region are computationally identified next to the hydrophobic CNT walls \cite{joseph2008carbon,ritos2014flow}.

\section{A universal nanofluidic slip flow model}
Considering the balance of the involved hydrodynamic forces in nanofluidics and the interfacial length scale, we define the \textit{slip flow number}. While the inertial to viscous forces as indicated by Reynolds number is the fundamental number representing hydrodynamic forces involved in the fluid flow behaviour, with scales approaching the nanometres, interfacial forces can also become important as the fluid slip at interfaces can be significant. The interfacial tension (although a static force) is the parameter associated with the viscosity at the interfacial fluid layer. The viscosity is also directly linked with the interfacial friction at the slip layer in nano-scales. The Capillary number which is the ratio between viscous to interfacial forces is then the potential representative number in these scales. Thus the ratio between Reynolds to Capillary numbers as the balance of all forces multiplied by a function of the interfacial layer scale factor, \ie $f(s/d)$ is used to define the slip flow number $\Omega$,
\begin{equation}
\Omega=\frac{Re}{Ca}\;f(\frac{s}{d})
=\frac{\rho u d/\mu}{\mu u/\gamma_i^{e\!f\!f}}\;f(\frac{s}{d})=\frac{\rho \gamma_i^{e\!f\!f} d}{\mu^2}\;f(\frac{s}{d})\label{eq:om-f(Lam)}
\end{equation}
where $\rho$ is the fluid (bulk) density, $u$ is the fluid (average) velocity, $d$ is the characteristic dimension of the system \eg tube diameter, $\mu$ is the dynamic viscosity for the bulk fluid, and $s$ is the interfacial fluid layer characteristic thickness ($s\approx0.5nm$). This number is based on the bulk values of density and viscosity but accounts for the sub-interface variations of these quantities intrinsically via the interfacial energy. When selecting the simulation scales, care has been taken to ensure that the validity of these considerations is not affected by a non-local viscosity \cite{todd2008nonlocal}. We then assume a simple power function for $f$ as $\big(\frac{s}{d}\big)^\alpha$. 

A large number of pressure-driven flow systems (infinitely long tubes and parallel plates with different diameters and distances) are simulated where interfacial forces are varied from weak to strong, attractive to repulsive modes (in total more than 660 simulation cases including 22 tube and channel sizes each simulated for more than 30 different interfacial force conditions). Analyses of the results show that when $\alpha=1.2$, all the $\Omega-\varepsilon$ curves (for all systems examined) converge into a unique behaviour (with the smallest sum of squared differences among the curves; see Appendix \ref{App:A}, section 4 for more details). Thus we propose the dimensionless \textit{slip flow number} in the following specific form,
\begin{equation}
\Omega=\frac{\rho\gamma_i^{e\!f\!f} s}{\mu^2}\;\big(\frac{s}{d}\big)^{0.2}\label{eq:om}.
\end{equation}

The characteristic length, $d$ is readily the inner diameter of circular tubes including the depletion thickness. For nanochannels, we implement a geometrical transformation of the parallel plates into analogous tubes with identical solid-fluid interface area and volume of fluid that leads to $d=2l$ where $l$ is the separation distance of the plates. With such transformation, consistent behaviour is found for slip flows in tubes and between parallel plates when characterized by the same definition of \textit{slip flow number}.

To examine the generality of the model, we also tested variations of the force models (with constant or exponentially decaying solid-fluid force functions with different decay lengths). Four different force functions were examined on 3 tube diameters, each simulated over more than 20 different force strengths (more than 240 cases in total). The analysis showed that the conclusions are universal with the same behaviour observed at similar \textit{slip flow numbers} (see Appendix \ref{App:A}, section 7 for more details). 

\begin{figure*}[!htbp]
	\centering
	\includegraphics[width=0.7\linewidth]{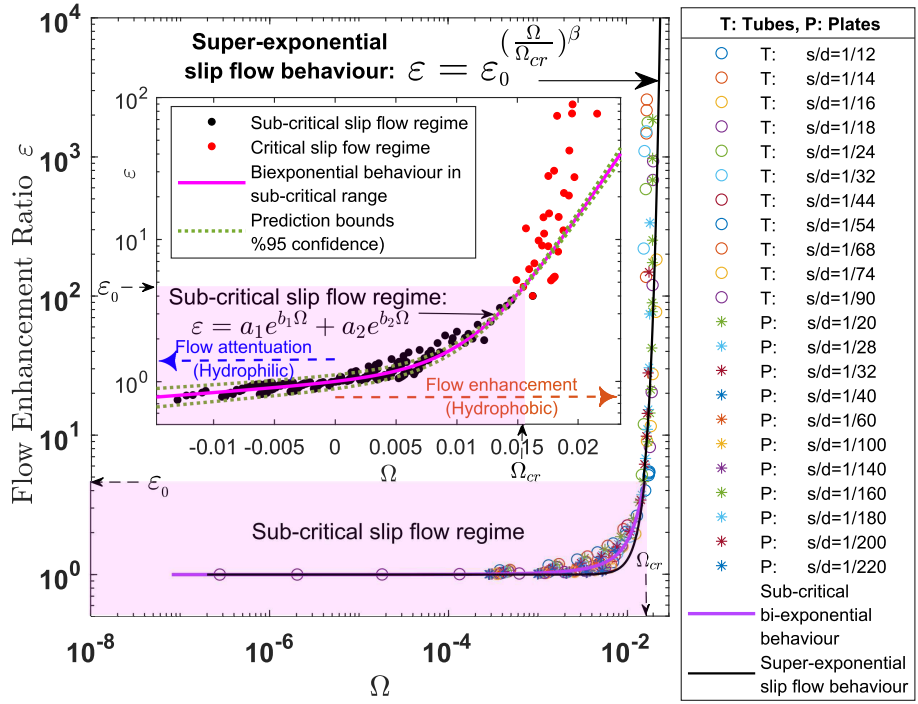}
	\caption {\textbf{The universal nanofluidic slip flow model.} The unified $\Omega-\varepsilon$ relationship provides a universal slip flow model for nanofluidics (obtained for tubes and parallel plates). The model is presented as a superexponential function (Eq. \ref{eq:SEL}: $\varepsilon=\varepsilon_0^{(\Omega/\Omega_{cr})^\beta}$) for the critical regime ($\Omega\!>\!\Omega_{cr}$) with the optimal coefficients' values: $\varepsilon_{_0}=4.50$, $\Omega_{cr}=0.0154$ and $\beta= 4.13$. A bi-exponential function can best describe the sub-critical regime corresponding to flow attenuation and enhancement due to hydrophilic and hydrophobic effects, respectively (the inset).}
	\label{fig:omega}
\end{figure*}

As shown in Fig. \ref{fig:omega}, our simulations reveal a sub-critical slip flow behaviour followed by a critical regime. The critical regime can be described by the following universal superexponential slip flow function: 
\begin{equation}
\varepsilon=\varepsilon_{_0}^{(\Omega/\Omega_{cr})^\beta}\label{eq:SEL}
\end{equation}
with  $\varepsilon_{_0}\approx4.50$, $\Omega_{cr}\approx0.0154$ and $\beta\!\approx\!4.13$. In this model, $\varepsilon_{_0}$ is the flow enhancement ratio corresponding to the threshold of critical behaviour (the critical \textit{slip flow number}, $\Omega_{cr}$) beyond which the slip flow turns into the critical superexponential regime, \ie approaching the limit of frictionless walls. 

For $\Omega\!<\!\Omega_{cr}$ (sub-critical), the slip flow regime is best characterized as a bi-exponential behaviour, with the following equation: \begin{equation}
\varepsilon = a_1e^{b_1\Omega} + a_2e^{b_2\Omega}
\end{equation} where the coefficients (with $\%95$ confidence bounds) are
$a_1 =      0.97$  ($0.962, 0.987$),
$b_1 =       15.14$  ($12.14, 18.15$),
$a_2 =     0.03$  ($0.024, 0.037$) and
$b_2 =         304$  ($290.6, 317.5$). The sub-critical regime describes both flow attenuation for hydrophilic ($\Omega<0$)  and flow enhancement for hydrophobic ($\Omega>0$) cases up to $\Omega\!\approx\!\Omega_{cr}$ (Fig. \ref{fig:omega}: inset)

\begin{figure*}[!htbp]
	\centering
	\includegraphics[width=0.39\linewidth]{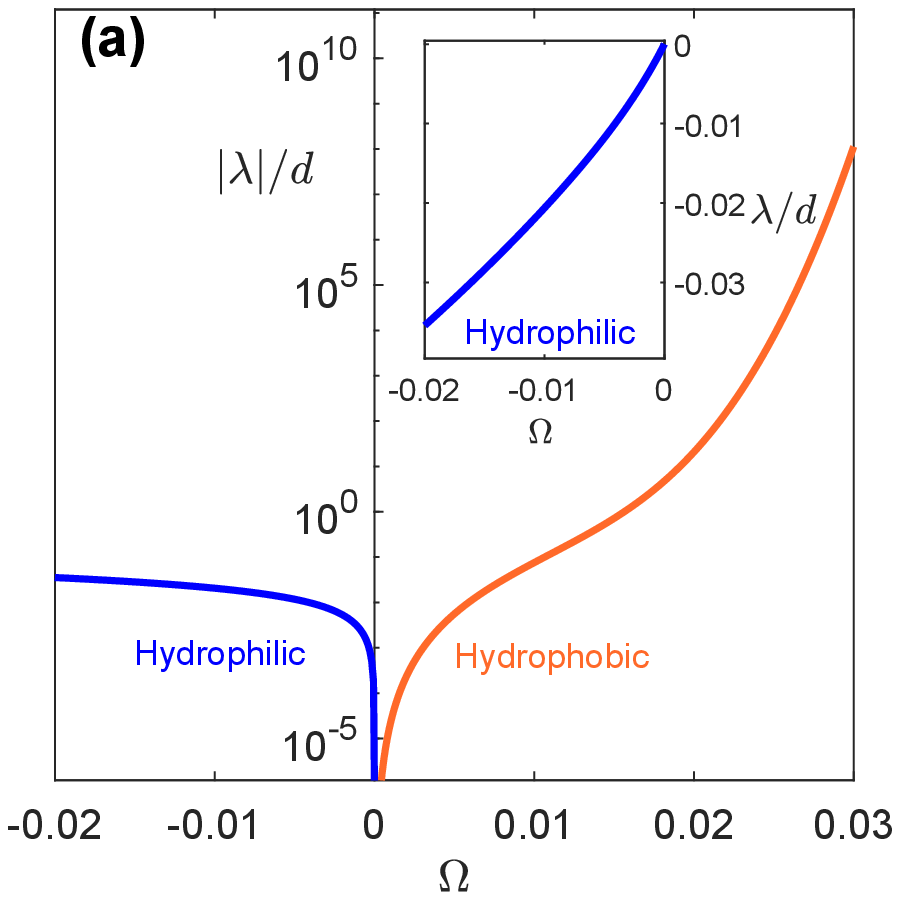}
	\includegraphics[width=0.39\linewidth]{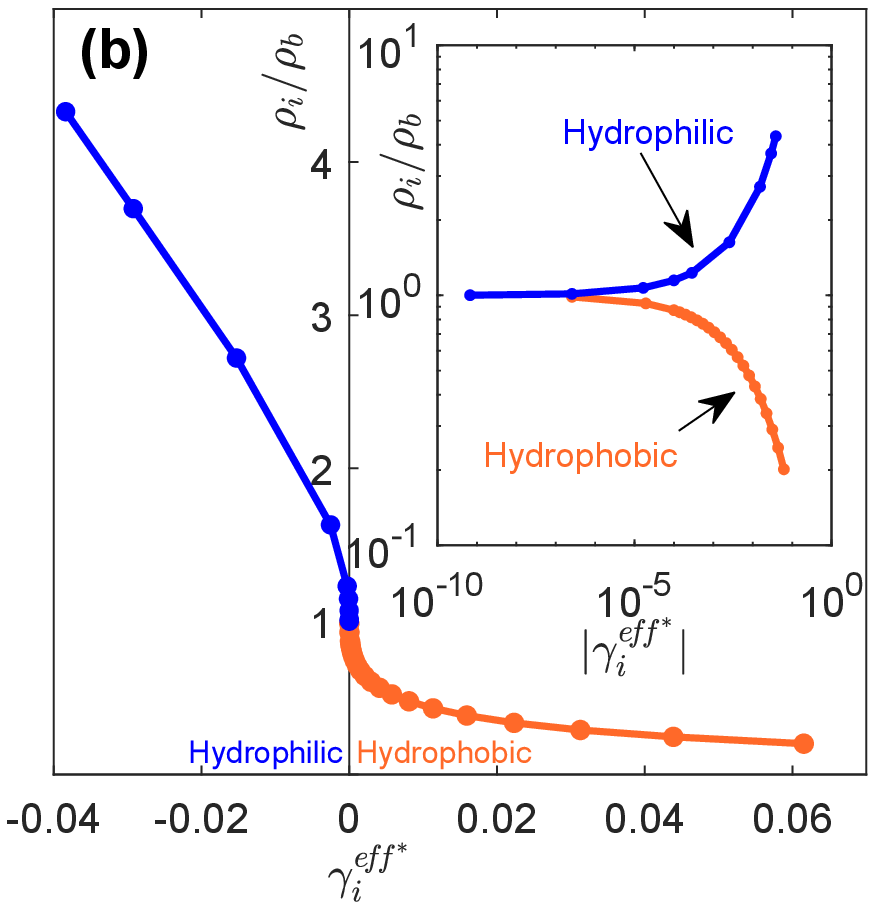}
	\caption {\textbf{The unified nanofluidic slip flow characteristics.} a) The normalized apparent slip length as a function of $\Omega$. Note the absolute value $|\lambda|$ in the graph. The inset shows the hydrophilic side of the graph with a normal scale for the vertical axis showing the actual negative values of $\lambda/d$. b) Normalized average fluid density at the interfacial layer ($\rho_i/\rho_b$) versus the dimensionless effective interfacial energy ($\gamma_i^{{e\!f\!f}^*}=\frac{\gamma_i^{e\!f\!f}}{\mu^2/(\rho s)}$). The inset in (b) shows the density variations with a logarithmic vertical axis. Note the absolute value $|\gamma_i^{{e\!f\!f}^*}|$ used in the inset.
	}\label{fig:lamda-rho}
\end{figure*}

The normalized apparent slip lengths can also be shown as a function of $\Omega$ via the analytical relationship for tubes, $\varepsilon=1+8\frac{\lambda}{d}$ (Fig. \ref{fig:lamda-rho}). The interfacial density, $\rho_i$, \ie the average fluid density in the interface region next to walls (with characteristic thickness of $s\!\approx\! 0.5nm$) is also shown to be a function of the nondimensionalized effective interfacial energy, $\gamma_i^{{e\!f\!f}^*}=\frac{\gamma_i^{e\!f\!f}}{\mu^2/(\rho s)}$ in Fig. \ref{fig:lamda-rho}. The value $\rho_i/\rho_b$ (where $\rho_b$ is the bulk density) indicates the range of fluid densification (and rarefaction) near hydrophilic (and hydrophobic) surfaces. A strong hydrophilicity (\eg $\gamma_i^{e\!f\!f}=-15\;mJ/m^2$) corresponding to $\gamma_i^{{e\!f\!f}^*}=$
$-0.0074$ can lead to water densification in the interface with $\rho_i/\rho_b\approx2.05$. This densification is supported by experimental evidences such as the nanoacoustic measurements of a $5$-times higher water density at Al$_2$O$_3$ hydrophilic surfaces in water (tending to bulk density in about $1nm$ distance) \cite{mante2014probing}. Contrarily, a typical  $\gamma_i=83\;mJ/m^2$ for hydrophobic CNTs \cite{van2017direct} gives $\gamma_i^{{e\!f\!f}^*}=0.041$, which corresponds to the rarefaction at the interface ($\rho_i/\rho_b\approx0.25$).

\section{Experimental validation}

\subsection{Experimental estimation of the interfacial energy, $\gamma_i$ for nanotube materials\label{gammai_values}}

Wetting properties of the nanotubes and their constituent materials were investigated in several previous studies as summarized in Table \ref{tab:gammai}. We select the following values as the interfacial energies of nanotube materials when calculating the $\Omega$: $\gamma_{i_{(CNT:graphene)}}=83mJ/m^2$ \cite{van2017direct}, $\gamma_{i_{(BNNT)}}=23mJ/m^2$ \cite{yum2006measurement}, $\gamma_{i_{(Polysulfone-UF)}}=48.93mJ/m^2$ \cite{ammar2017nanoclay}, $\gamma_{i_{(Polycarbonate)}}=38.0mJ/m^2$ \cite{bhurke2007surface}. Particularly, we select the value of $\gamma_{i_{(CNT:graphene)}}$ from Reference \cite{van2017direct} in which the interfacial energy of graphene is directly measured using a surface force apparatus independently instead of an indirect estimation from other parameters (contact angle). The role of energy perturbations or functionalization effects are discussed later (Fig. \ref{fig:eps-d-validation}).

\subsection{Model validation}
\begin{figure*}[t]
	\centering
	\includegraphics[width=0.7\linewidth]{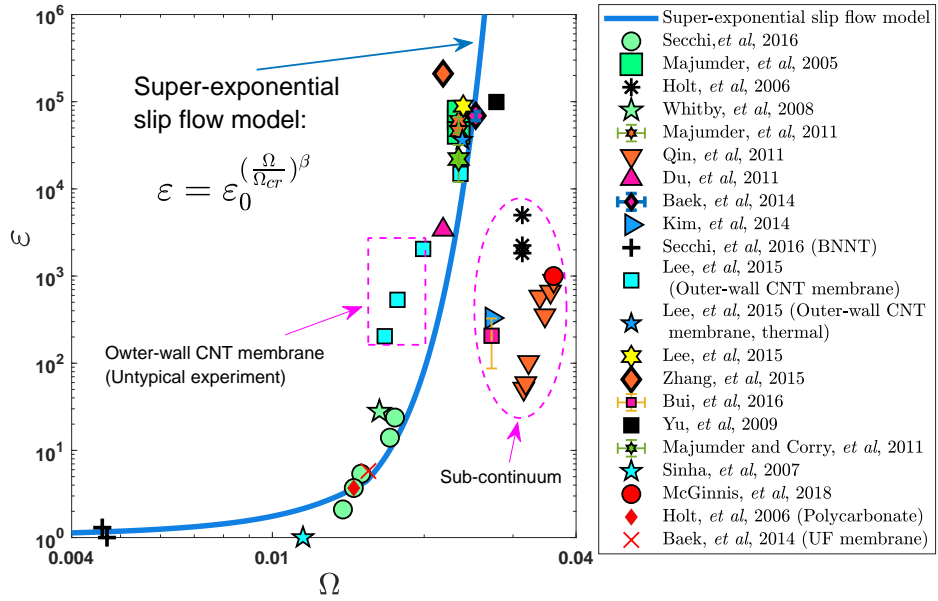}
	\caption {\textbf{The universal nanofluidic slip flow model versus experimental data.} Data extracted from 40 experiments on water flow in nanotubes are shown in comparison with the proposed universal slip flow model (see Table \ref{tab:exp-lit-data} for experimental details). Note that the model is proposed for continuum slip flow regime and the sub-continuum data are shown for comparison. Also note untypical conditions in outer-wall membranes where the voids between vertical close-ended tubes were used as flow channels. Also see Fig. \ref{fig:eps-d-validation}.}
	\label{fig:eps-om-validation}
\end{figure*}

We compared our model with a large number of experimental data to evaluate the validity of the revealed behaviour (Fig. \ref{fig:eps-om-validation}). The model is proposed for the continuum flow regime up to the limit of the validity of classical hydrodynamics which can be up to a channel width of about ten molecular diameters, \ie $d\!>\!3nm$ \cite{travis1997departure,noy2007nanofluidics}. The sub-continuum experiments indicate an evidently different behaviour. We note that in the experiments conducted on outer-wall membranes by Lee, \etal 2015 \cite{lee2015carbon}, the flow channels have been the voids between outer walls of vertical CNTs, not inside tubes. These untypical conditions may explain larger differences between these data and the model predictions. The details of experiments are summarized in Table \ref{tab:exp-lit-data}. 

\begin{figure*}[!htbp]
	\centering
	\includegraphics[width=\linewidth]{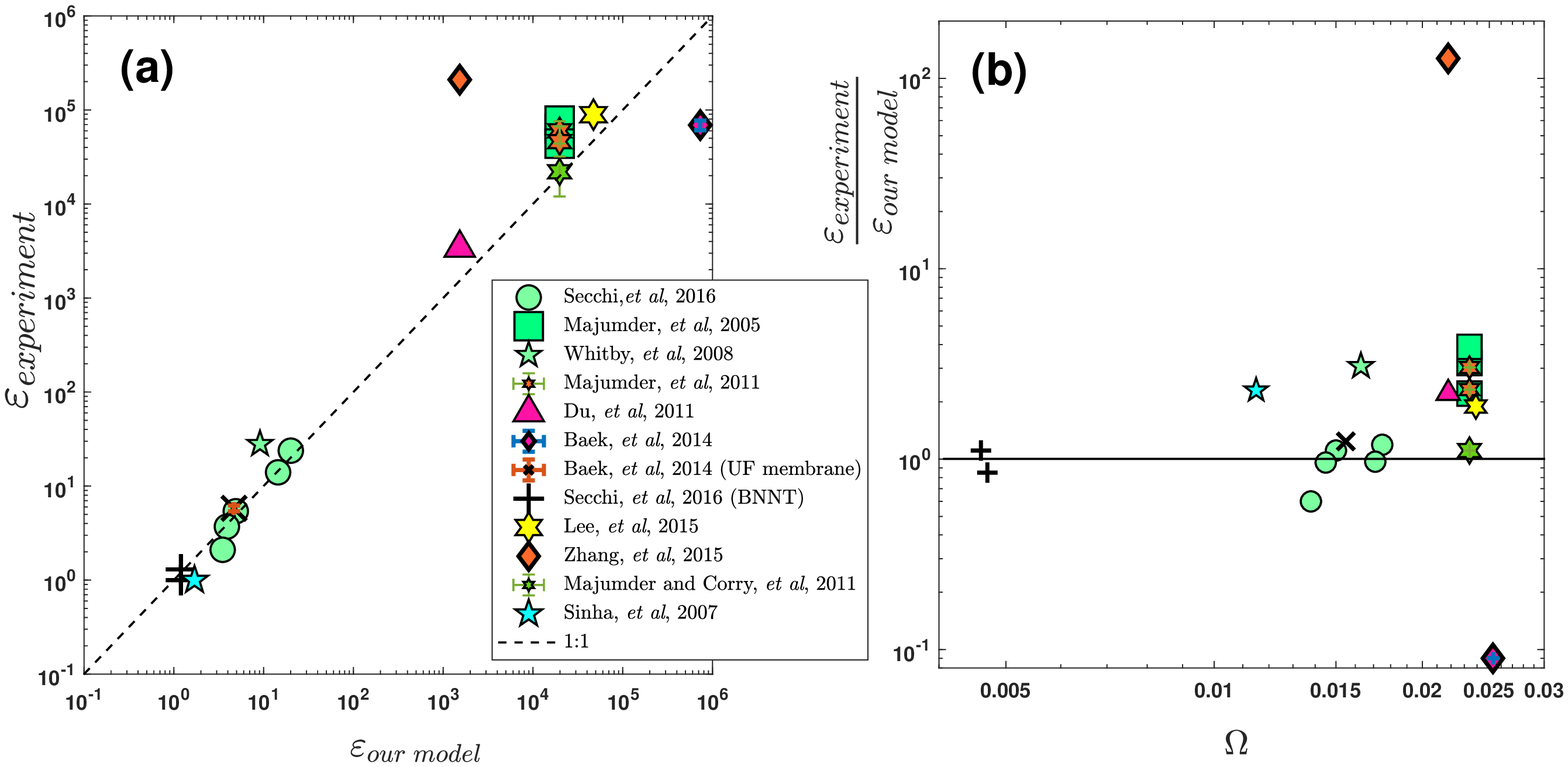}
	\caption {\textbf{Model predictions versus experimental data.} a) The model predictions of flow rate enhancement ratios, $\varepsilon$ are shown versus the experimental values for continuum flow experiments in CNTs (20 experiments). Untypical experiments and those affected by sub-continuum flow ($d\lessapprox3$ nm) are excluded. b) The ratio between experimental measurements of $\varepsilon$ and corresponding predictions of our model is shown versus the \textit{slip flow number}, $\Omega$. This ratio approaches unity for the majority of data. The experiment conducted by Zhang \etal 2015 \cite{zhang2015preparation} that most deviates from our predictions can be highly affected by entrance effects with $\mathcal{R}_p/\mathcal{R}_c=10.31$}
	\label{fig:eps-om-validation-second}
\end{figure*}

A comparison between our predicted enhancement ratios and experimental values as shown in Fig. \ref{fig:eps-om-validation-second}-a demonstrates a close agreement between the model and experiments across the entire range of $\varepsilon$ variations. The ratio between experimental $\varepsilon$ and corresponding predicted ones approaches unity for the majority of continuum experiments (Fig \ref{fig:eps-om-validation-second}-b).

\begin{table*}[]
	\centering
	\caption{\textbf{Experimental data from literature on water flow in nanomembranes versus our predictions.} Here, for the calculation of $\Omega=\frac{\rho \gamma_i s}{\mu^2}(\frac{s}{d})^{0.2}$, the following parameters are assigned: $\rho=997kg/m^3$, $s=0.5nm$, $\mu=1.002mPa.s$, $\gamma_{i_{(CNT:graphene)}}=83mJ/m^2$ \cite{van2017direct}, $\gamma_{i_{(BNNT)}}=23mJ/m^2$ \cite{yum2006measurement}, $\gamma_{i_{(Polysulfone-UF)}}=48.93mJ/m^2$ \cite{ammar2017nanoclay}, and $\gamma_{i_{(Polycarbonate)}}=38.0mJ/m^2$ \cite{bhurke2007surface}. The values are selected based on the review summary provided in Table \ref{tab:gammai}.}
	\label{tab:exp-lit-data}
	\resizebox{\textwidth}{!}{%
		\begin{tabular}{|c|c|c|c|c|c|c|c|c|c|c|}
			\hline
			No. & Ref.  & Material &  d (nm) &\begin{tabular}[c]{@{}c@{}} CNT length,\\ L ($\mu m$)\end{tabular}& $\varepsilon$, Experiment & \begin{tabular}[c]{@{}c@{}}Slip length,\\    $\lambda$ (nm)\end{tabular} & $\Omega$ &	$\varepsilon_{our\;model}$	&	$\frac{\varepsilon_{\;experiment}}{\varepsilon_{\;our\;model}}$	&	$\mathcal{R}_p/\mathcal{R}_c$
			\\ \hline
			1 & Holt, \etal, 2006  \cite{holt2006fast} 			  		  & CNT & 1.65 			&2	& 5000.0 & 1031.0 & 0.0325 &	Sub-continuum	&	-	&	2.43
			\\ \hline
			2 & Holt, \etal, 2006  \cite{holt2006fast}					  & CNT & 1.65 			&3	& 2240.0 & 461.8 & 0.0325 &	Sub-continuum	&	-	&	0.73
			\\ \hline
			3 & Holt, \etal, 2006  \cite{holt2006fast}					 & CNT & 1.65 			&2.8	& 1830.0 & 377.2 & 0.0325 &	Sub-continuum	&	-	&	0.64
			\\ \hline
			4 & Holt, \etal, 2006  \cite{holt2006fast}					& Polycarbonate & 2.10   &6	& 3.7 & 0.7 & 0.0145  & Sub-continuum	&	-	&	0.00
			\\ \hline
			5 & Majumder, \etal, 2005   \cite{majumder2005nanoscale}    & CNT & 7				& 34& 63333.3 & 55415.8 & 0.0243 &	19807.4	&	3.20	&	7.68
			\\ \hline
			6 & Majumder, \etal, 2005   \cite{majumder2005nanoscale}     & CNT & 7				&34 & 77017.5 & 67389.5 & 0.0243 			&	19807.4	&	3.89	&	9.34
			\\ \hline
			7 & Majumder, \etal, 2005   \cite{majumder2005nanoscale}     & CNT & 7				 &126 & 43859.6    & 38376.3 & 0.0243 	&	19807.4	&	2.21	&	1.44\\ \hline

			8 & Secchi, \etal, 2016 \cite{secchi2016massive}             & CNT & 30             &  0.7   & 23.8 & 85.3 & 0.0182 &	20.1	&	1.19	&	0.60
			
			\\ \hline

			9 & Secchi, \etal, 2016 \cite{secchi2016massive}			 & CNT & 34 			 &0.45 	& 14.0 & 55.2 & 0.0177 			&	14.5	&	0.97	&	0.62
			\\ \hline

			10 & Secchi, \etal, 2016 \cite{secchi2016massive}			  & CNT & 66            & 0.9   & 5.4 & 36.1 & 0.0156 
			&	4.9	&	1.11	&	0.23
			\\ \hline

			11 & Secchi, \etal, 2016 \cite{secchi2016massive}			 & CNT & 76            &0.8  & 3.7 & 25.3 & 0.0150 
			&	3.9	&	0.96	&	0.20
			\\ \hline

			12 & Secchi, \etal, 2016 \cite{secchi2016massive}			 & CNT & 100           & 1   & 2.1 & 14.0 & 0.0143 
			&	3.5	&	0.60	&	0.12
			\\ \hline

			13 & Secchi, \etal, 2016 \cite{secchi2016massive} 			 & BNNT & 46            &  0.6& 1.0 & 0.0 & 0.0049 
			&	1.2	&	0.85	&	0.05
			\\ \hline

			14 & Secchi, \etal, 2016 \cite{secchi2016massive} 		      & BNNT & 52            & 0.7 & 1.3 & 2.0 & 0.0048 
			&	1.2	&	1.11	&	0.06\\ \hline

			15 & Whitby, \etal, 2008 \cite{whitby2008enhanced}           & CNT & 43$\pm$3          &78 & 28$\pm$6 & 145.1$\pm$32.3 & 0.0169 
			&	9.1	&	3.08	&	0.01
			\\ \hline

			16 & Majumder and Corry, 2011  \cite{majumder2011anomalous}  & CNT & 7 				&100& 22000 $\pm$10000 & 19249$\pm$8750 & 0.0243 
			&	19807.4	&	1.11	&	0.91
			\\ \hline

			17 & Majumder, \etal, 2011   \cite{majumder2011mass}  & CNT & 7					 &34-126& 60000$\pm$16000 & 52499$\pm$14000 & 0.0243

			&	19807.4	&	3.03	&	1.96
			
			\\ \hline

			18 & Majumder, \etal, 2011   \cite{majumder2011mass}  & CNT & 7					 &81 & 46000$\pm$21000 & 40249$\pm$18375 & 0.0243 
			
			&	19807.4	&	2.32	&	2.34
			
			\\ \hline

			19 & Majumder, \etal, 2011   \cite{majumder2011mass}  & \begin{tabular}{@{}c@{}}CNT: tip-functionalized \\with a polypeptide spacer\\ and the anionically charged\\ dye molecule.\end{tabular} & 7 										&34-126 & 200$\pm$30 & 174$\pm$26 & -

			&	No info on $\gamma_i$	&	-	&	0.01

			\\ \hline

			20 & Majumder, \etal, 2011   \cite{majumder2011mass} & \begin{tabular}{@{}c@{}}CNT: tip and core\\-functionalized \\with a polypeptide spacer\\ and the anionically charged\\ dye molecule.\end{tabular}  & 7 						&34-126& $<$5.3 & 3.8 & -

			&	No info on $\gamma_i$	&	-	&	0.00
			
			\\ \hline

			21 & Qin, \etal, 2011 \cite{qin2011measurement}  & CNT & 1.59 					& 805 & 51.0 & 9.9 & 0.0327

			&	Sub-continuum	&	-	&	0.00
			
			\\ \hline

			22 & Qin, \etal, 2011  \cite{qin2011measurement}  & CNT & 1.52					& 805 & 59.0 & 11.0 & 0.0330

			&	Sub-continuum	&	-	&	0.00
			\\ \hline

			23 & Qin, \etal, 2011  \cite{qin2011measurement}  & CNT & 1.42 				& 805 & 103.0 & 18.1 & 0.0334

			&	Sub-continuum	&	-	&	0.00
			\\ \hline

			24 & Qin, \etal, 2011  \cite{qin2011measurement}  & CNT & 1.10 				& 805 & 580.0 & 79.6 & 0.0352 
			
			&	Sub-continuum	&	-	&	0.00
			
			\\ \hline

			25 & Qin, \etal, 2011  \cite{qin2011measurement}  & CNT & 0.98 				& 805 & 354.0 & 43.2 & 0.0360

			&	Sub-continuum	&	-	&	0.00
			\\ \hline

			26 & Qin, \etal, 2011  \cite{qin2011measurement}  & CNT & 0.87 				& 805 & 662.0 & 71.9 & 0.0369 
			
			&	Sub-continuum	&	-	&	0.00
			
			\\ \hline

			27 & Qin, \etal, 2011  \cite{qin2011measurement}  & CNT & 0.81					& 805 & 882.0 & 89.2 & 0.0375 
			
			&	Sub-continuum	&	-	&	0.00
			
			\\ \hline

			28 & Sinha, et al, 2007 \cite{sinha2007induction} & CNT & 250.00 				&10   & 1.0 & 0.0 & 0.0119 
			
			&	2.3	&	0.44	&	0.01
			
			\\ \hline

			29 & Du, \etal, 2011 \cite{du2011membranes}  & MWCNT & 10						&4000 & 3403.0 & 4252.5 & 0.0226 
			
			&	1529.9	&	2.22	&	0.01
			
			\\ \hline

			30 & Baek, \etal, 2014  \cite{baek2014high}  & CNT & 4.8$\pm$0.9 				&200 & 69039$\pm$8122 & 41422.8$\pm$4873 & 0.0262 
			
			&	730782.1	&	0.09	&	0.98
			\\ \hline

			31 & Baek, \etal, 2014  \cite{baek2014high}  & Ultrafilteration (UF) membrane & 5.7$\pm$2.5 		& 0.1 & 5.8$\pm$0.4 & 3.4$\pm$0.3 & 0.0155 
			&	4.7	&	1.24	&	0.19
			
			\\ \hline

			32 & Kim, \etal, 2014   \cite{kim2014fabrication}  & CNT & 3.3$\pm$0.7 		&20-50& 330.0 & 135.7 & 0.0282
			
			&	Sub-continuum	&	-	&	0.02
			
			\\ \hline

			33 & Lee, \etal, 2015  \cite{lee2015carbon} & \begin{tabular}[c]{@{}c@{}}Flow in voids between CNTs\\ - Outer wall membrane\end{tabular} & 37.80 		&1300 & 203.9 & 958.9 & 0.0173 
			
			&	11.4	&\begin{tabular}[c]{@{}c@{}}17.92\\ Untypical experiment\end{tabular}		&	0.00
			
			\\ \hline

			34 & Lee, \etal, 2015  \cite{lee2015carbon} & \begin{tabular}[c]{@{}c@{}}Flow in voids between CNTs\\ - Outer wall membrane\end{tabular} & 28.80 		&1300	& 533.7 & 1917.8 & 0.0184 
			
			&	23.0	&\begin{tabular}[c]{@{}c@{}}23.17\\ Untypical experiment\end{tabular}		&	0.01
			
			\\ \hline

			35 & Lee, \etal, 2015  \cite{lee2015carbon}  & \begin{tabular}[c]{@{}c@{}}Flow in voids between CNTs\\ - Outer wall membrane\end{tabular} & 15.80 		&1300	& 2047.1 & 4041.1 & 0.0206 
			
			&	148.6	&	\begin{tabular}[c]{@{}c@{}}13.78\\ Untypical experiment\end{tabular}	&	0.01
			
			\\ \hline

			36 & Lee, \etal, 2015  \cite{lee2015carbon} & \begin{tabular}[c]{@{}c@{}}Flow in voids between CNTs\\ - Outer wall membrane\end{tabular} & 6.70 			&1300	& 14885.5 & 12465.8 & 0.0245 
			
			&	27846.8	&\begin{tabular}[c]{@{}c@{}}0.53\\ Untypical experiment\end{tabular}		&	0.05
			
			\\ \hline

			37 & Lee, \etal, 2015  \cite{lee2015carbon}  & \begin{tabular}[c]{@{}c@{}}Flow in voids between CNTs\\Thermally treated CNT\\ - Outer wall membrane\end{tabular} & 6.50														&1300 & 36247.2 & 29450.0 & 0.0247 
			
			&	39496.1	&\begin{tabular}[c]{@{}c@{}}0.92\\ Untypical experiment\end{tabular}		&	0.11
			
			\\ \hline

			38 & Lee, \etal, 2015  \cite{lee2015carbon} & CNT open ended (inner flow) & 6.40 			&200	& 89126.0 & 71300.0 & 0.0248

			&	47195.4	&	1.89	&	1.68

			\\ \hline

			39 & Zhang, \etal, 2015 \cite{zhang2015preparation}  & CNT & 10 						&120	& 210000.0 & 262498.8 & 0.0226

			&	1529.9	&	137.27	&	10.31
			
			\\ \hline

			40 & Bui, \etal, 2016 \cite{bui2016ultrabreathable}  & CNT & 3.30 					& 23000	& 206.5$\pm$119.5 & 84.8$\pm$49.3 & 0.0282 
			
			&	Sub-continuum	&	-	&	0.00
			\\ \hline

			41 & Yu, \etal, 2009  \cite{yu2008high}  & CNT & 3 									&750	& 99210.0 & 37203.4 & 0.0288 
			
			&	Sub-continuum	&	-	&	0.23
			\\ \hline

			42 & McGinnis, et al, 2018 \cite{mcginnis2018large}  & CNT & 0.67-1.27 				&0.5-1.5	& 1000 & 83-158 & 0.0361 
			&	Sub-continuum	&	-	&	0.57
			\\ \hline
		\end{tabular}%
	}
\end{table*}

The model features the onset of the critical regime ($\Omega_{cr}$) where flow enhancement intensifies superexponentially with $\Omega$ (Fig. \ref{fig:omega}). Both the critical regime's onset and the growth rates are evidently in agreement with the experimental data (Fig. \ref{fig:eps-om-validation}).

\begin{figure*}[!htbp]
	\centering
	\includegraphics[width=0.45\linewidth]{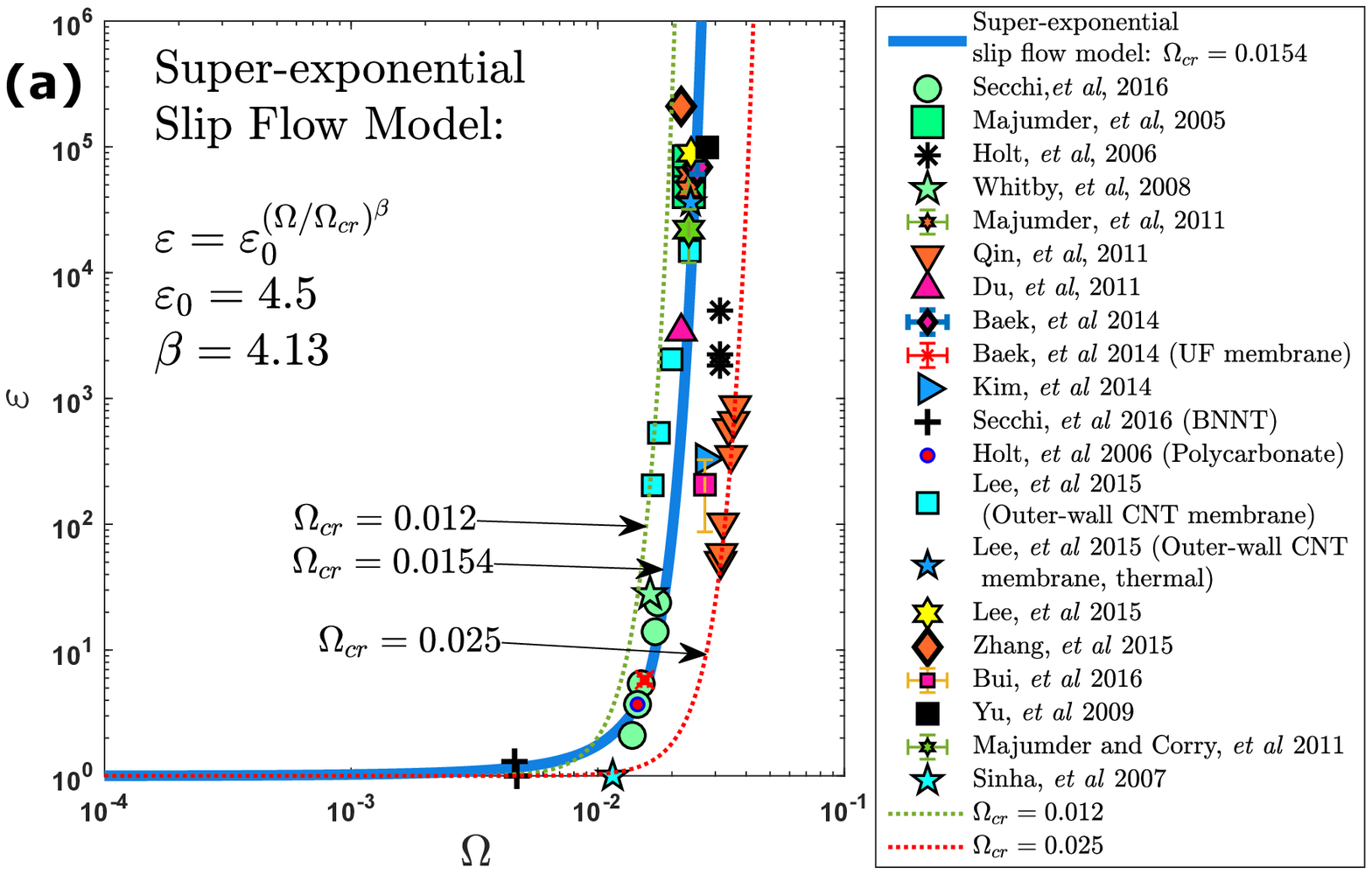}
		\includegraphics[width=0.3\linewidth]{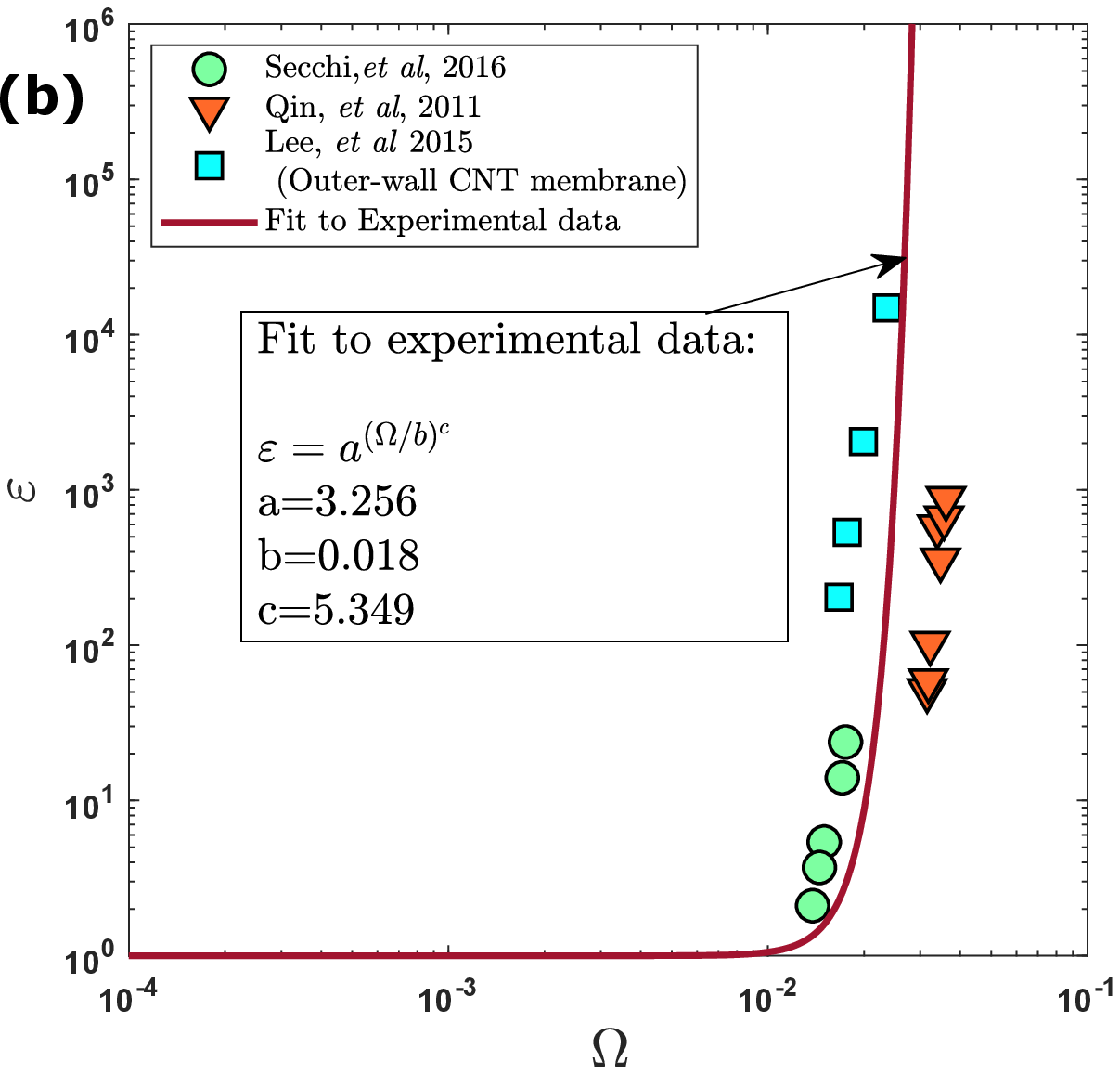}
	\caption {\textbf{Independent experimental derivation of model parameters ($\Omega_{cr}$ and $\beta$)}. a) The analysis of experimental data suggests the range of $0.012<\!\Omega_{cr}\!<\!0.025$ by which the proposed superexponential model predicts $\%95$ of the experimental data from $41$ experiments. The computationally-calculated value of $\Omega_{cr}=0.0154$ lies within this range. b) An equation is fitted ($\varepsilon=a^{(\Omega/b)^c}$) to data from selected experiments in which all test conditions except the tube diameters have been constant. The analysis suggests a superexponential growth factor $c=5.349$, close to the computationally-derived value of $\beta=4.13$ (in $\varepsilon=\varepsilon_0^{(\Omega/\Omega_{cr})^\beta}$).  
	}
	\label{fig:eps-om-validation-om-cr-range}

\end{figure*}

To further examine the validity of this model, we independently derived the main features of the model from experiments \ie a) the onset of the critical regime ($\Omega_{cr}$) and b) the superexponential growth rate. Firstly, we showed that a range of $0.012<\!\Omega_{cr}\!<\!0.025$ for the proposed superexponential model ($\varepsilon=\varepsilon_0^{(\Omega/\Omega_{cr})^\beta}$, $\varepsilon_0=4.5$, $\beta=4.13$) can cover $\%95$ of experimental data from $41$ experiments including the limiting sub-continuum data. The computationally-determined value of $\Omega_{cr}=0.0154$ developed for continuum regime lies within the given range derived from experimental data (see Fig. \    \ref{fig:eps-om-validation-om-cr-range}-a). Secondly, we independently investigate the growth rate of $\varepsilon$ in experiments. For this purpose, we select those reports in which multiple tests were performed on nanotubes of different diameters. Thus, $\Omega$ in each set of these experiments varies with only diameters (with other materials and test conditions being constant). Therefore the growth rate of $\varepsilon$ versus $\Omega$ in those experiments is a pure experimental outcome independent from any other assumption. As illustrated in Fig. \ref{fig:eps-om-validation-om-cr-range}-b, this analysis further validates the proposed superexponential growth model with the experimentally-fitted value of $\beta$ equal to $5.349$ close to the computationally-derived value of $4.13$ for superexponential coefficient in our model ($\beta$ in $\varepsilon=\varepsilon_0^{(\Omega/\Omega_{cr})^\beta}$).

\section{Discussion}

\subsection{Entrance effects}
The transition from a macroscopic reservoir to a nano-scale pore with streamlines being bent while entering the channel is a source of viscous dissipation. Considering a nanopore of radius $R$ from which the flow enters a channel (tube) of the same radius, we can assume that the total hydrodynamic resistance $\mathcal{R}_{pc}$ is the sum of the resistances through pore and channel, \ie $\mathcal{R}_{pc}=\mathcal{R}_p+\mathcal{R}_c$. The total pressure drop $\Delta P$ is also the sum of pressure drops through pore and channel, $\Delta P= \Delta P_p+\Delta P_c$.
Also flow of channel is $Q_c=\Delta P_c/\mathcal{R}_c$ while the flow through pore is $Qp=\Delta P_p/\mathcal{R}_p$. The entrance-corrected flow is then $Q_{pc}={Q_p=Q_c}=\Delta P/ (\mathcal{R}_c+\mathcal{R}_p)$. 

For a nanopore, the flow rate can be written as
\begin{equation}
Q_p=\frac{R^3}{3\mu}\Delta P_p.
\end{equation}
The flow rate through a pore is not significantly affected by slip conditions since the dissipation source is mainly geometric.

For the no-slip boundary conditions, it can be shown that entrance effects are apparently negligible for tube lengths exceeding a few channel radii \cite{kavokine2021fluids}. 

Considering a non-zero slip length $\lambda$, the flow rate through a tube (channel) is given as
\begin{equation}
Q_c=\frac{\pi R^4}{8\mu L}\bigg(1+\frac{4\lambda}{R}\bigg)\Delta P_c
\end{equation}
Therefore the entrance-corrected flow rate in case of slip is
\begin{equation}
Q_{pc}=\frac{R^3}{3\mu}\frac{\Delta P}{1+\frac{8L}{3\pi(R+4\lambda)}}.
\end{equation}
From the above equation it can be seen that when $\lambda> R$,  the hydrodynamic resistance is dominated by entrance effects as long as $\lambda>L$. 

\begin{figure}[h]
	\centering
	\includegraphics[width=\linewidth]{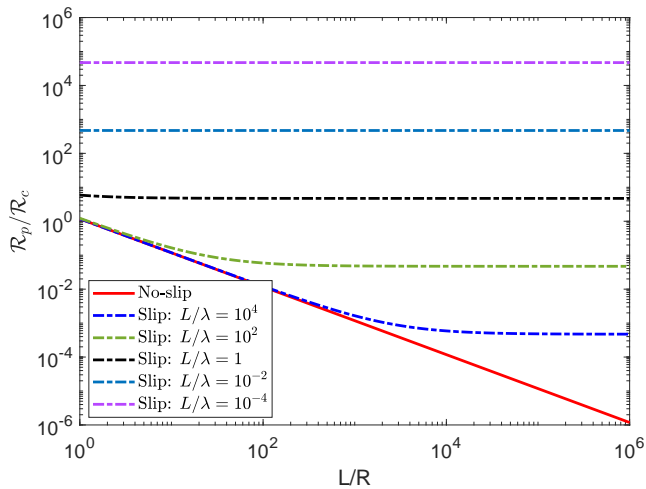}
	
	\caption {\textbf{Criteria for the significance of entrance effects.}. The ratio between the pore (entrance) hydrodynamic resistance and channel (tube) resistance, $\mathcal{R}_p/\mathcal{R}_c$ versus $L/R$ where $L$ and $R$ are the channel length and radius, respectively. The viscous entrance effects are dominant when $\lambda>L$.
	}
	\label{fig:RpRc}
\end{figure}

 We use the value of $\mathcal{R}_p/\mathcal{R}_c=3\pi(R/L+4\lambda/L)/8$ as a criterion for the significance of entrance effects. Fig. \ref{fig:RpRc} shows this criterion in which the limit of $\mathcal{R}_p/\mathcal{R}_c>1$ can be defined as the threshold of the flow being dominated by viscous entrance effects. With slip length $\lambda$ reaching the order of channel length $L$, flow behaviour approaches such a threshold \cite{kavokine2021fluids}. 

 Our model which is based on channel slip flow in the continuum regime is limited to relatively long tubes ($L/\lambda>1$) in which the entrance effects are negligible. When comparing experimental data from the literature to our model, we show if the entrance effects are significant or negligible (Table \ref{tab:exp-lit-data}). In most cases, the entrance effects are shown to be negligible. In a few cases, the entrance resistance can be effective on the flow rates for some factors up to at most a factor of 10. However, in those experiments, enhancement factors have been several orders of magnitude suggesting that the model predictions based on channel flow could yet be practically reasonable, compared to the entrance effects which are effective for only some factors.

 In particular, Sisan and Lichter, 2011 \cite{sisan2011end} showed that the experiments reported by Majumder \etal 2005 \cite{majumder2005nanoscale} can be up to 9 times faster than physically allowed values for frictionless tubes as they should be dominated by entrance effects. We also show that $\mathcal{R}_p/\mathcal{R}_c$ for those experiments could be between 1.44 and 9.34 (Table \ref{tab:exp-lit-data}). The findings of Ref. \cite{sisan2011end} are also consistent with our work as the flow rates reported by Mjumder et al. 2005 are 2.21 – 3.8 times faster than our model predictions. Also the experiment reported by Zhang \etal \cite{zhang2015preparation} that most deviates from our predictions (Fig. \ref{fig:eps-om-validation-second}-b) should be highly affected by entrance effects with $\mathcal{R}_p/\mathcal{R}_c=10.31$.

\subsection{Continuum nanofluidic slip flow as iso-$\gamma_{i}$ model behaviour}
\begin{figure*}[!htbp]
	\centering
	\includegraphics[width=0.49\linewidth]{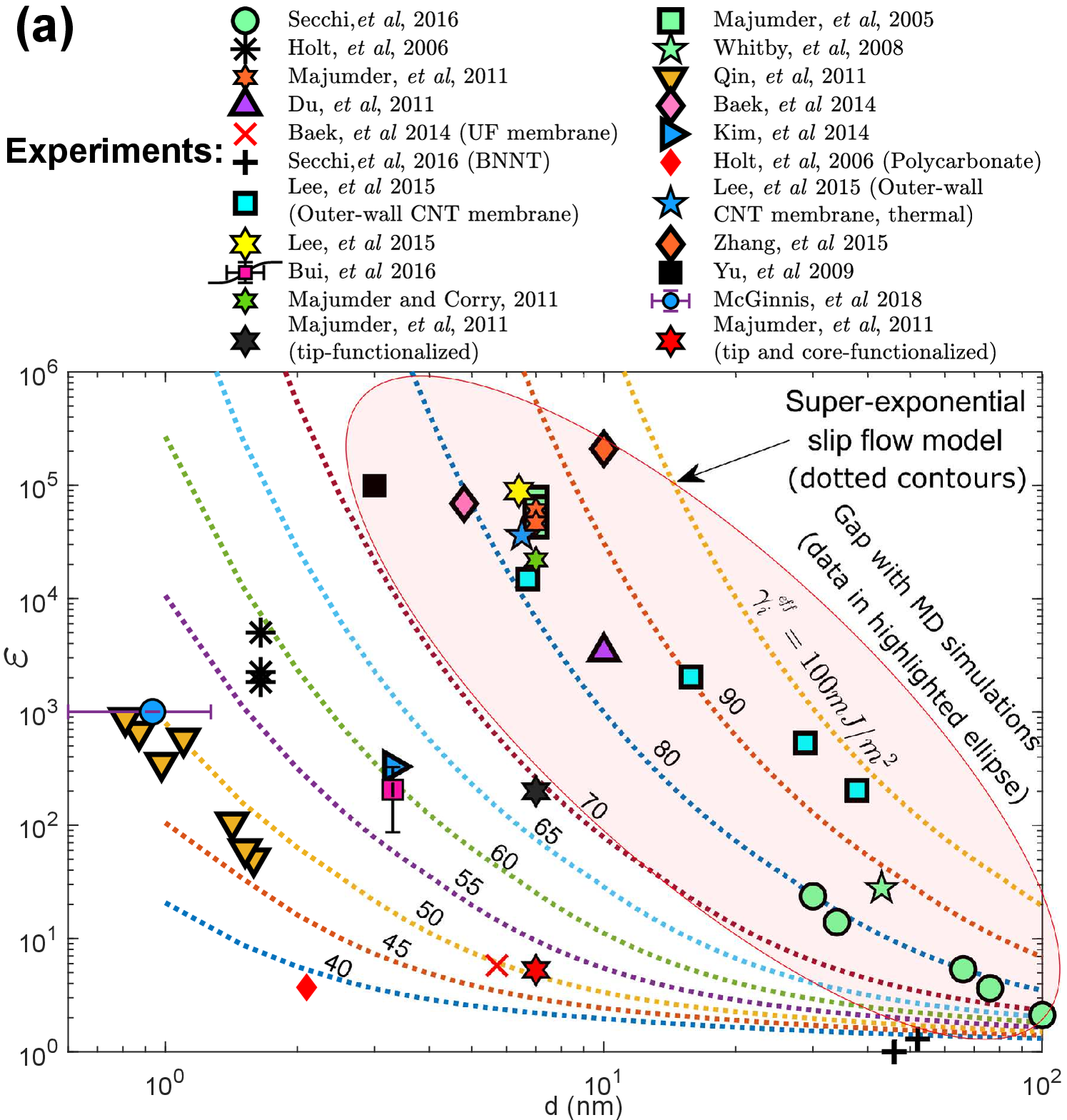}
	\includegraphics[width=0.49\linewidth]{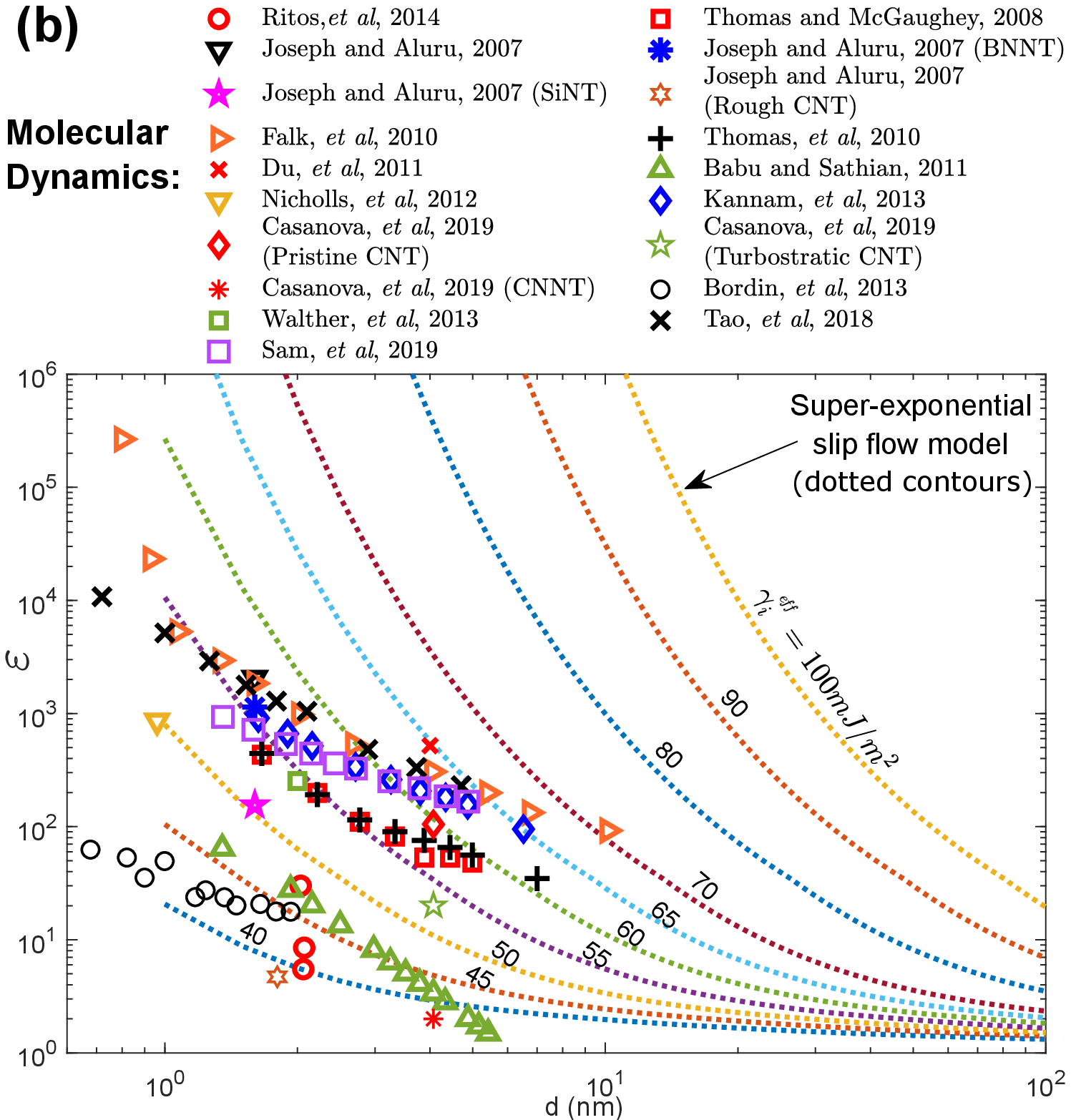}
	\caption {\textbf{Literature data for nanotube-water systems versus the universal (superexponential) slip flow model.} Original data as ($d,\varepsilon$) values from $42$ experiments (a) and $97$ MD simulations (b) are shown (see Appendix \ref{App:C} for the details of MD data). Literature data points are for CNTs, unless mentioned otherwise. Our model is presented as dotted iso-$\gamma_i^{{e\!f\!f}}$ lines in both a and b as the solution of Eq. \ref{eq:SEL} \ie $\varepsilon = f(\Omega)=f(d, \gamma_{i}, constants)$ in which for $\Omega$ (Eq. \ref{eq:om}), the following parameters are assumed as constants: $\rho=997kg/m^3$, $s=0.5nm$, and $\mu=1.002mPa.s$. A large portion of experimentally reported ($d,\varepsilon$) values (data in the highlighted ellipse in a) have never been replicated by MD simulations (b). The cause of this significant gap is discussed in the text. Despite 6 orders of magnitude variations, the majority of experimental data  in continuum slip regime ($d>3nm$) shown in a can be explained well with our iso-$\gamma_i^{{e\!f\!f}}$ model with graphene's value of $\gamma_{i}\!\!=\!\!83\pm7$ $mJ/m^2$ \cite{van2017direct}. Sub-continuum experimental data ($d<3nm$) appear to behave differently.}
	\label{fig:eps-d-validation}
\end{figure*}

In Fig. \ref{fig:eps-d-validation} we show our slip flow model (shown as  $\gamma_i^{{e\!f\!f}}$ contours) versus original data from experiments (Fig. \ref{fig:eps-d-validation}-a) and MD simulations (Fig. \ref{fig:eps-d-validation}-b) \cite{sinha2007induction,kim2014fabrication,bui2016ultrabreathable,yu2008high,mcginnis2018large,babu2011role,nicholls2012water,kannam2013fast,bordin2013relation,walther2013barriers,tao2018confinement,sam2019water}. The original data are reported values of $\varepsilon$ versus reported tube diameters, $d$. Thus the data points on Fig. \ref{fig:eps-d-validation} are pure literature data (with no assumptions) showed against the solution of our model (iso-$\gamma^{e\!f\!f}_i$ lines). The model can predict almost whole range of experimental $\varepsilon$ variations for CNTs in continuum regime ($d>3nm$) over 6 orders of magnitude (Fig. \ref{fig:eps-d-validation}-a) with finite variations of interface energies around the value of graphene ($\gamma_{i_(graphene)}\!\!=\!\!83\pm7$ $mJ/m^2$ \cite{van2017direct}). The flow mainly dominated by sub-continuum regime ($d<3nm$) appears to behave differently as the experimental data deviate from our model's iso-$\gamma^{e\!f\!f}_i$ contours at graphene's range. 

In particular, Secchi, \etal 2016 \cite{secchi2016massive} and Lee, \etal 2015 \cite{lee2011performance} have reported multiple experiments with constant test conditions except tube diameters (thus constant interfacial energies). These data lie within similar iso-$\gamma_i^{{e\!f\!f}}$ lines predicted by our model in Fig. \ref{fig:eps-d-validation}-a further evidencing the validity of  predictions.

Of particular importance could be the similar iso-$\gamma_i$ lines attributed to the data of Secchi, \etal 2016 \cite{secchi2016massive} (one of the most accurate measurements to date) and most of the other continuum flow experiments as shown in Fig. \ref{fig:eps-d-validation}-a. Similar $\gamma_{i}$ levels indicating equivalent behaviours again suggest the feasibility of large $\varepsilon$ in nanotubes with larger diameters, supporting many experiments \cite{lee2015carbon,whitby2008enhanced,baek2014high,zhang2015preparation}.

\subsection{functionalization effects}
The role of energy levels can also describe the major impacts of functionalizations on flow enhancements \cite{casanova2018surface,joseph2008carbon}. For instance, chemical hydrophilic functionalization of 7-nm CNT membranes is reported by Majumder \etal 2011 to reduce $\varepsilon$ for more than 4 orders of magnitude \cite{majumder2011mass}. According to our model, a reduced $\gamma_{i}$\ from $\approx83$ (orange hexagrams in Fig. \ref{fig:eps-d-validation}-a),  to $\approx70mJ/m^2$ (black hexagram) and $\approx50mJ/m^2$ (red hexagram) corresponds to original, tip and core-functionalized nanotubes in this report, respectively.

\subsection{Differences between CNTs and BNNTs}
Insight into the surprising dissimilarity of CNTs and BNNTs (Secchi, \etal 2016 \cite{secchi2016massive}) could also be provided considering $\gamma_i{_{BNNT}}=21-26\;mJ/m^2$ \cite{yum2006measurement} being lower than $\gamma_i{_{CNT}}=83\;mJ/m^2$ \cite{van2017direct}. Such differences in  $\gamma_i$ match closely with the predictions of our model (see Fig. \ref{fig:eps-d-validation}-a). This rationale can complete the \textit{ab initio} findings \cite{tocci2014friction} that inked the larger friction on boron nitride compared to graphene to a greater corrugation of free energy \cite{tocci2014friction}, but suggested only a 3-fold friction difference between CNTs and BNNTs which was much smaller than experimental observations. 
Indeed, experiments revealed the differences between CNTs and BNNTs \cite{secchi2016massive} in much greater scales where CNTs enhanced flow up to a factor of 100, but similarly-sized BNNTs produced no flow enhancements.
These differences convinced the authors of Ref. \cite{secchi2016massive} to consider the corrugation effects as minor mechanisms and suggest mechanisms beyond these effects to describe the phenomenon. Now our findings here may further explain the actual scales of the differences between CNTs and BNNTs by potential differences of their interfacial energies.

\subsection{The gap between experimental data and MD simulations}
The proposed model sheds light on the significant gap between MD predictions and experimental data (large  $\varepsilon$ values reported in experiments for $d\!>\!5nm$, \ie data in the highlighted ellipse in Fig. \ref{fig:eps-d-validation}-a). The insufficiency of MD empirical force models to retrieve all hydrophobicity features including beyond vdW effects and hydrogen bonding networks might be the reason for such a gap. Furthermore, when attempting to recover the interfacial (\eg carbon-water) interactions in nanotubes, MD investigators have normally relied on the tuning of contact angles \cite{werder2003water}. However, reproduction of the solid-liquid energy states as reference values rather than $\theta$ could be crucial in characterizing nonbounded interactions of water on graphene (see Ref. \cite{leroy2015parametrizing} for a detailed study). Thus, slight inaccuracies in intermolecular energy parameters used in MD simulations could also be the origin of significant deviations
from experiments.

\subsection{$\varepsilon-\theta$ relationship}
\begin{figure*}[!htbp]
	\centering
	\includegraphics[width=0.65\linewidth]{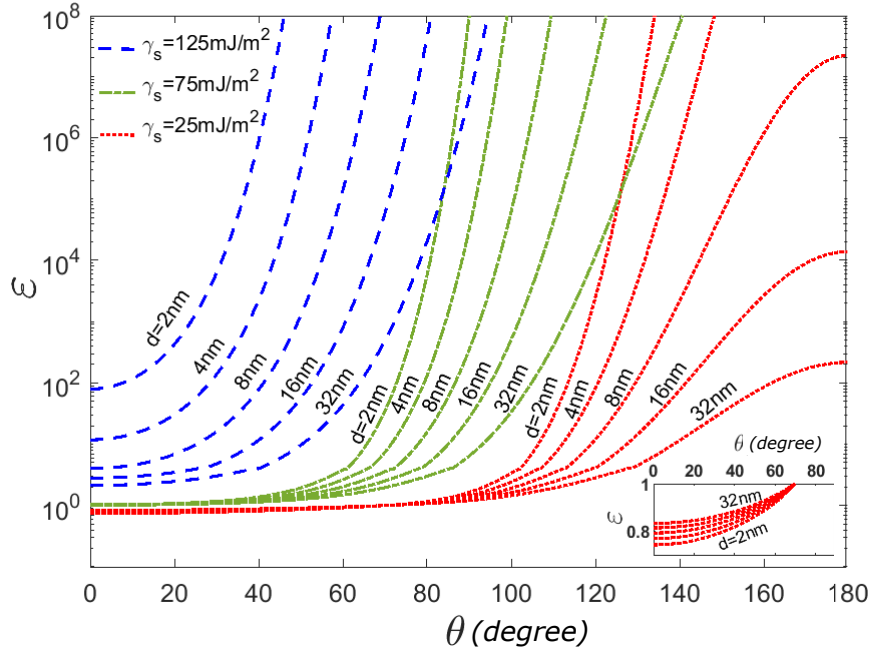}
	\caption {\textbf{Flow enhancement ratio versus contact angle.} The predicted $\varepsilon$ depends on the tube diameter $d$, surface energy $\gamma_s$ and contact angles $\theta$ (via Young's equation \ref{eq:young}). The inset magnifies the hydrophilic portion ($\varepsilon\!<\!1$). Dependence of the flow enhancement on the energy state -instead of a sole contact angle- explains how moderately hydrophobic CNTs (with reported $\theta_{graphene}\approx70^{\circ}-90^{\circ}$ \cite{van2017direct,rafiee2012wetting,kozbial2014study}, and $\theta_{CNT}\approx70^{\circ}$ \cite{ma2010dispersion}) could induce extraordinary flow enhancements. The model differs from the few existing theoretical models that attribute typical flow enhancements observed in nanotubes to very large contact angles ($\theta >150^\circ$) \cite{wu2017wettability}. In calculating $\varepsilon$ (Eq. \ref{eq:SEL}) as the function of $\Omega$ (Eq. \ref{eq:om}), the following parameters are assumed as constants: $\rho=997kg/m^3$, $s=0.5nm$, and $\mu=1.002mPa.s$.}
	\label{fig:theta-eps}
\end{figure*}

Our model accounts for a combined role of contact angles, $\theta$ and surface energy, $\gamma_s$ in flow enhancement behaviours, as shown in Fig. \ref{fig:theta-eps}. The graph illustrates how the flow slippage (enhancement) does not solely depend on $\theta$; hence even traditionally-defined hydrophilic surfaces (with $\theta\!<\!90^{\circ}$) can cause flow slippage depending on  $\gamma_s$. Here the difference between a conventional definition of hydrophobicity ($\theta\!>\!90^{\circ}$) and the principle of positive interfacial energies used here is highlighted. Thus the findings can also explain the “unexpected” slip on surfaces with $\theta\!<\!90^{\circ}$ observed in experiments \cite{bonaccurso2002hydrodynamic,lee2012water} and simulations \cite{ho2011liquid}.

\section{Conclusion}

An continuum approach coupled with interfacial force models in simulating solid-fluid intermolecular effects enabled us to provide an inclusive picture of continuum slip flow regimes in nanofluidics. We showed that the interfacial energies that originate the interfacial fluid viscosity and density variations can be responsible for ultra-fast flows in nanotubes. The slip mechanism attributes to the variations of nanotube diameters and interfacial energies. We developed a universal model that quantifies flow enhancement ratios of nanotubes and nano-channels as a superexponential function of a single dimensionless \textit{slip flow number}, $\Omega$. The model could explain all previously measured flow enhancements filling a significant gap between experimental data and MD results. The model features a critical threshold ($\Omega_{cr}$) beyond which the flow enhancement ratios ($\varepsilon$) turn from a normal exponential to a superexponential growth. Insights were provided into long-standing puzzling observations in nanofluidic systems such as scattering of reported $\varepsilon$ data by six orders of magnitude, four orders of magnitude decrease of $\varepsilon$ due to hydrophilic functionalizations \cite{majumder2011mass} and slippage of water on low-contact-angle surfaces \cite{bonaccurso2002hydrodynamic,ho2011liquid}. We highlighted a significant gap between previous MD simulations and experimental results where the ultrafast flow rates observed in wider-than-5nm tubes have never been computationally predicted by MD simulations (three-order-of-magnitude differences in $\varepsilon$). Our study suggests the role of actual interfacial energy levels as a clue to addressing such a gap. 

We limited our model to relatively long nanotubes (when entrance effects are negligible) in continuum flow regimes ($d>3nm$). The entrance effects are only significant when the slip lengths exceed the channel length. An analysis of the experimental data showed that the model predictions can be reasonably accurate for the majority of previous experiments.

Our model can reduce the complicated problem of functionalization effects on nanoscale permeability to a fundamental or experimental determination of the altered interfacial energy. This advancement could aid tailoring of separation membranes in which both permeability and selectivity of nanotubes are tunable \cite{tunuguntla2017enhanced,majumder2011mass} by reversible functionalization techniques \cite{kohler2019water,wang2010reversible}, promising broad applications in water purification, ion-exchange and energy conversion.

\begin{acknowledgments}
This research is funded by the Australian Research
Council via the Discovery Projects (No. DPI120102188 and No. DP140100490).
The authors also acknowledge support from the
National Natural Science Foundation of China (Grant
No. 51421006). The authors thank Prof. Billy Todd for useful discussions.
\end{acknowledgments}


\appendix
\section{Computational details\label{App:A}}

\subsection{Continuum-based simulation in nanofluidics}
Despite the discrete nature of molecules in nanofluidic systems, a continuum-based approach has proved valid in describing the transport phenomena at the nanoscale \cite{sparreboom2009principles,thomas2009water,wu2017wettability}. The molecular dynamics and continuum hydrodynamics are shown to be different only in scales of a few layers of molecules \cite{succi2007lattice,qiao2003ion,sparreboom2010transport}. As suggested by MD studies, for water nanofluidic systems, the limit of continuum transport mechanism (where the flow can be modelled using continuum-based relations) can be up to a channel width of about ten molecular diameters, \ie $d\!>\!3nm$ \cite{travis1997departure,noy2007nanofluidics}. The characteristic length (tube diameter) of $1.66nm$ is also proposed below which the transition to the sub-continuum mechanism takes place \cite{thomas2009water,thomas2008reassessing,wu2017wettability}. Thus, the majority of experiments of nanotube flow systems can be studied using a continuum approach, provided that the interfacial features are properly captured. Lattice-Boltzmann is shown valid to investigate the nano-scale local hydrodynamic interactions, for example in modelling lipid systems (with a grid spacing of $\approx 3\AA$ being sufficiently accurate) \cite{brandner2019modelling,sterpone2015protein,chiricotto2016hydrodynamic}. In our continuum framework, we adopted a lattice-Boltzmann approach with intermolecular potential for interfaces to account for the interfacial interactions averaged on the scales of the interface.

\subsection{Lattice-Boltzmann method with intermolecular potentials at interfaces} 

The Boltzmann equation reads
\begin{eqnarray}
\frac{\partial f}{\partial t}+\boldsymbol{e} \cdot \boldsymbol{\nabla} f+\boldsymbol{F}\cdot \boldsymbol{\nabla}_e f = \Omega_{coll}
\end{eqnarray}
where the collision term $\Omega_{coll}=\frac{f^{eq}-f}{\tau}$ is the BGK model, $f\equiv f(\boldsymbol{x},t)$ is the particle distribution function in the phase space $(\boldsymbol{x},\boldsymbol{e})$ with $t$ being the time, $\boldsymbol{e}$ is the local molecular velocity, $\boldsymbol{F}$ is the external force experienced by each fluid particle, $f^{eq}$ is an equilibrium distribution function and $\tau$ is the characteristic relaxation time.

The Navier-Stokes equations are shown to be recovered with the equilibrium distribution function given by \cite{he1997lattice}
\begin{equation}
f_i^{eq}=\omega_i\rho\left(1+\frac{3(\boldsymbol{e_i}\cdot \boldsymbol{u})}{C^2}+\frac{9(\boldsymbol{ e_i}\cdot\boldsymbol{u})^2}{2C^4}-\frac{3 \boldsymbol{u}^2}{2C^2}\right)
\label{eq:feq}
\end{equation}
where $C=\delta_x/\delta_t$ is the characteristic lattice velocity with $\delta_x$ being the time step and $\delta_x$, the grid spacing. The weight factor $\omega_i$ is also assigned for each direction, the sum of which would be unity (see \cite{aidun2010lattice} for weight factors of different schemes). The study is performed using a D3Q19 scheme, while the parametric analyses on numerical variables (\eg relaxation time) ensured the generality of the results \cite{aminpour2018slip}.

The interparticle potential can be incorporated into the lattice-Boltzmann equation using a mean-filed approximation \cite{he1998discrete}. Considering $f^{eq}$ being the leading part of the distribution $f$, with the assumption of $\nabla_e f^{eq}\approx\nabla_e f$, the Boltzmann equation can be discritized in time and integrated leading to the following implicit expression \cite{porter2012multicomponent}
\begin{equation}
\begin{multlined}
f_i(\boldsymbol{x}+\boldsymbol{e_i}\delta_t,t+\delta_t)
-f_i(\boldsymbol{x},t)\\=\Omega_{coll}
+\frac{\delta_t}{2}\bigg[f_i^F(\boldsymbol{x}+\boldsymbol{ e_i}\delta_t,t+\delta_t)+f_i^F(\boldsymbol{x},t)\bigg]
\end{multlined}
\label{eq:fiefs}
\end{equation}
where $f_i^F$ is the forcing term defined as \cite{he1998novel}
\begin{equation}
f_i^F=\frac{\boldsymbol{F}\cdot(\boldsymbol{e_i}-\boldsymbol{u})}{\rho C^2}f_i^{eq}.
\end{equation}

With a transformation defined as $\bar{f_i}=f_i-\frac{\delta t}{2}f_i^F$ applied to Eq. \ref{eq:fiefs}, we obtain an explicit solution for the Boltzmann equation incorporated with interparticle forces 
\begin{equation}
\begin{multlined}
\bar{f_i}(\boldsymbol{x}+\boldsymbol{e_i} \delta t,t+\delta t)-\bar{f_i}(\boldsymbol{x},t)
=\bar{\Omega}_{coll}+\delta t f_i^F
\end{multlined}
\end{equation}
where $\bar{\Omega}_{coll}=\frac{\bar{f}_i^{eq}(\boldsymbol{x},t)-\bar{f}_i(\boldsymbol{x},t)}{\tau}$ and $\bar{f}_i^{eq}=[1-\frac{\boldsymbol{F}\cdot(\boldsymbol{e_i}-\boldsymbol{u})}{2\rho C^2}\delta t]f_i^{eq}$.

The method implements a mean-field interparticle potential that can provide a semi-mesoscopic solution to the interfacial effects. Such molecular interactions that appear as interfacial energies averaged in scales of several molecular layers are treated in this method using solid-fluid attractive or repulsive forces (with different force models examined). The values of interfacial energies are calculated using the integration of the difference in the normal and transverse pressure tensor components obtained in LB simulations assuring the accuracy of the method for the level of fluid confinements.

\subsection{Calculation of the interfacial energy in LBM simulations}

\begin{figure}[!htbp]
	\centering
	\includegraphics[width=\linewidth]{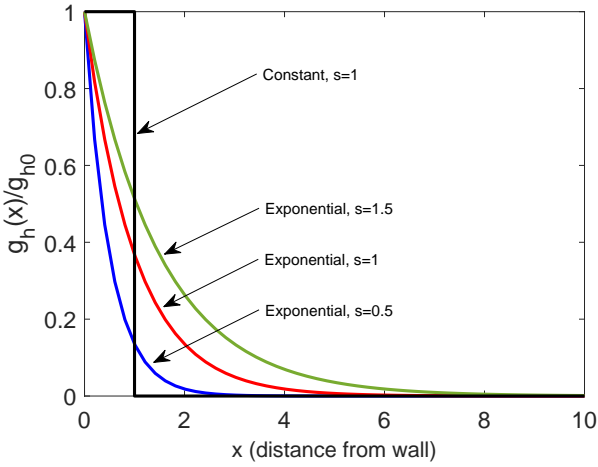}
	\caption {Intermolecular solid-fluid force models implemented in this study, \ie constant ($\boldsymbol{F_h}(x)=-\rho(x)\boldsymbol{{g}_{h_0}}$ for $0\!<\!x\!<\!s$ and $\boldsymbol{F_h}(x)=0$ for $s\!<\!x$) and exponential decay ($\boldsymbol{F_h}(x)=-\rho(x)\boldsymbol{{g}_{h}}(x)$ where $\boldsymbol{{g}_{h}}(x)=\boldsymbol{{g}_{h_0}}e^{-x/s}$). }\label{fig:force_functions}
\end{figure}

The pressure tensor $\veC P$ defined by $\veC\nabla \cdot \veC P=\veC \nabla (c_s^2\rho)-\veC F$ where $\veC F$ is the local boundary (interfacial) force acting on the fluid (when setting $\Delta t=1$) is equal to \cite{rowlinson2013molecular,lycett2015improved2}
\begin{equation}
\begin{multlined}
\veC P_{\alpha\beta}=\bigg(c_s^2\rho+\frac{c_s^2 g_h}{2}\psi^2+\frac{c_s^4 g_h}{4}(\veC \nabla \psi)^2+\frac{c_s^4 g_h}{2}\psi \Delta \psi \bigg)\delta_{\alpha \beta}\\
-\frac{c_s^4 g_h}{2}(\partial_\alpha \psi)(\partial_\beta \psi).
\label{eq:SCpressure}
\end{multlined}
\end{equation}
The variable $g_h$ specifies the intensity of the boundary (interfacial) force. This force is implemented in this study as a constant, $g_h(x)=g_{h_0}$  or an exponentially decaying function of $x$, $g_h(x)=g_{h_0} e^{(-x/s)}$ with different values for $s$ where $x$ is the distance from boundaries (see Fig. \ref{fig:force_functions}). The pseudopotential $\psi$ is given as $\psi(x)=\rho(x)$. We calculate the solid-fluid interfacial energy for an assumed planar interface at $x=0$ using the pressure tensor in (\ref{eq:SCpressure}) \cite{rowlinson2013molecular,lycett2015improved2}
\begin{equation}
\gamma_i=\int_{0}^{L}(P_{xx}-P_{yy})dx=-\frac{c_s^4 }{2}\int_{0}^{L}g_h\bigg(\frac{d\psi}{dx}\bigg)^2 dx. \label{eq:gamma_i}
\end{equation} 
The length $L$ is the length on which the boundary (interfacial) force is effective. 

\begin{figure}[h]
	\centering
	\includegraphics[width=0.5\linewidth,angle =-90]{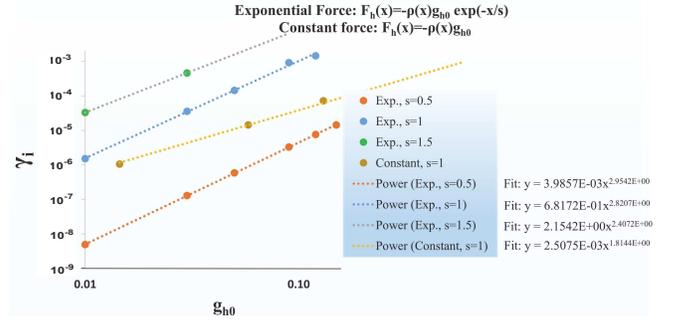}
	\caption {Non-linear relationship between the solid-fluid force intensity coefficient $g_{h_0}$ and the resulting interfacial energy $\gamma_i$ (in lattice units) for different intermolecular solid-fluid force models (constant or exponentially-decaying forces). For each force model (constant and exponentially decaying), we have changed the force factors (force intensity coefficient $g_{h_0}$ and the decay length $s$). $\gamma_i$ is calculated through a cell-wise integration of the pressure tensor as per Eq. (\ref{eq:gamma_i}). The power functions are fitted to each set of simulation data with a specific force function and $s$.  \label{fig:gh-gamma_i}.}
\end{figure}

We calculate the solid-liquid interfacial energy with equation (\ref{eq:gamma_i}) for each simulation (with a cell-wise integration) where the boundary force is implemented. It is worth mentioning that the relation between the force coefficient $g_h$ and the resulting interfacial energy is nonlinear as demonstrated in Fig. \ref{fig:gh-gamma_i}.

\subsection{Determination of the factor $\alpha$ in the definition of $\Omega$}

We defined a dimensionless \textit{slip flow number} $\Omega$ as follows:

\begin{equation}
\Omega=\frac{\rho \gamma_i^{e\!f\!f} d}{\mu^2}\;\big(\frac{s}{d}\big)^\alpha.
\label{eq:om-SI}
\end{equation}
To determine the coefficient $\alpha$ in Eq. \ref{eq:om-SI}, we implemented a least-square fitting analysis on the sub-critical portion of the results ($\varepsilon$ versus $\Omega$) for all simulations including tubes (with different diameters) and parallel plates (with various spacing) and variable intensities of interfacial energies.

\begin{figure}[!htbp]
	\centering
	\subfigure{\includegraphics[width=0.49\linewidth]{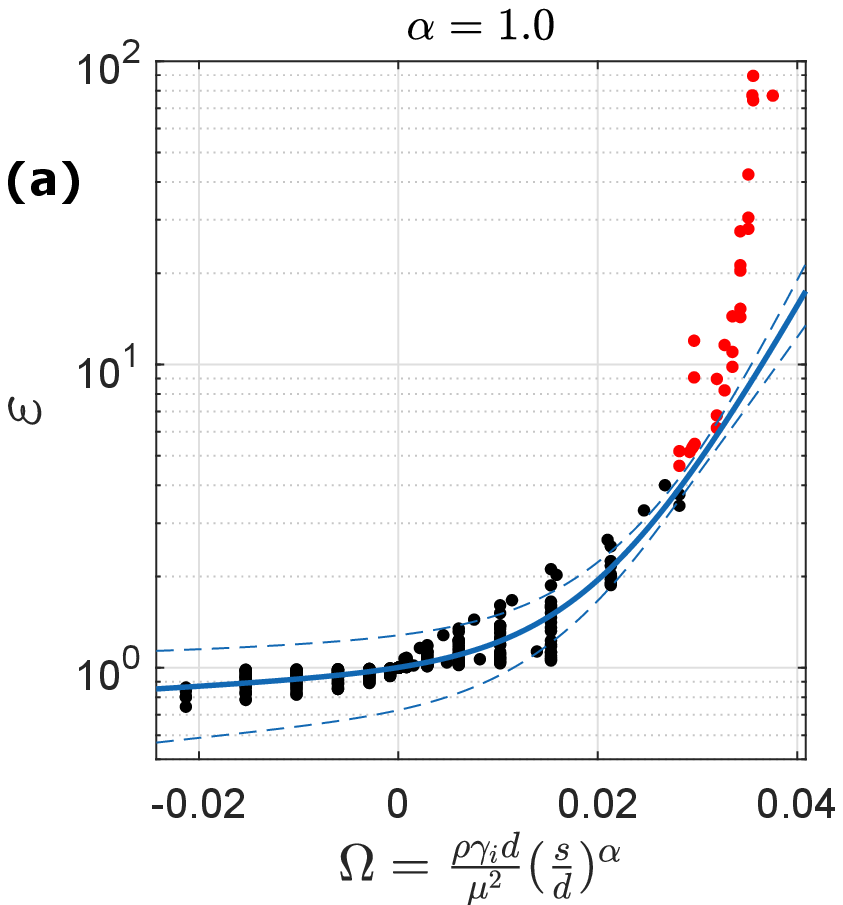}}
	\subfigure{\includegraphics[width=0.49\linewidth]{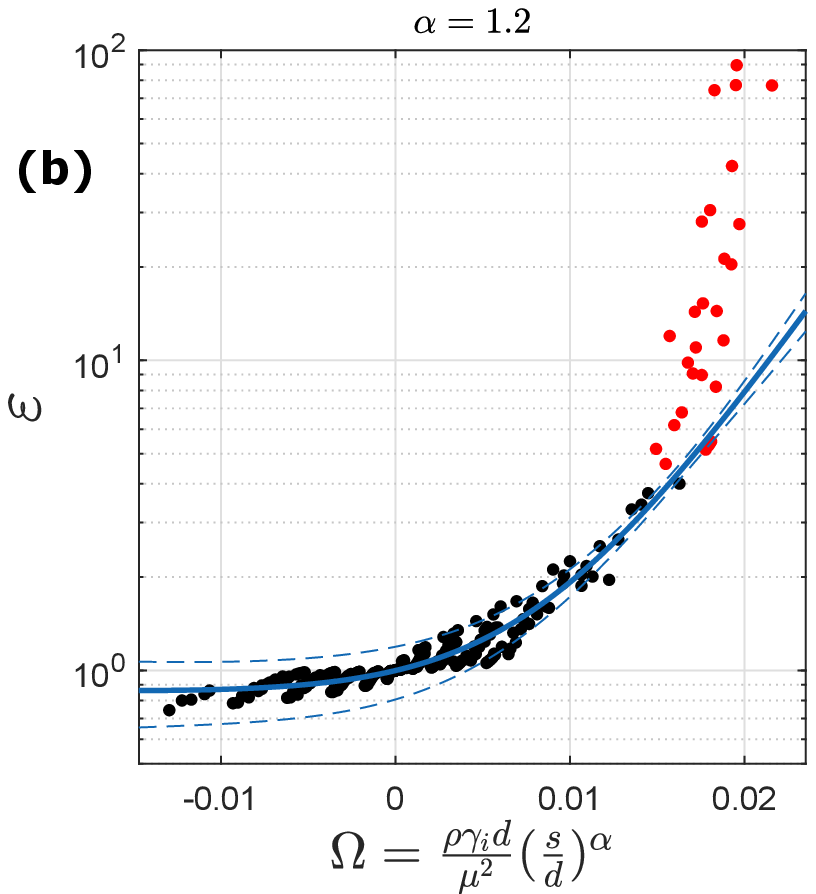}}\\
	\subfigure{\includegraphics[width=0.49\linewidth]{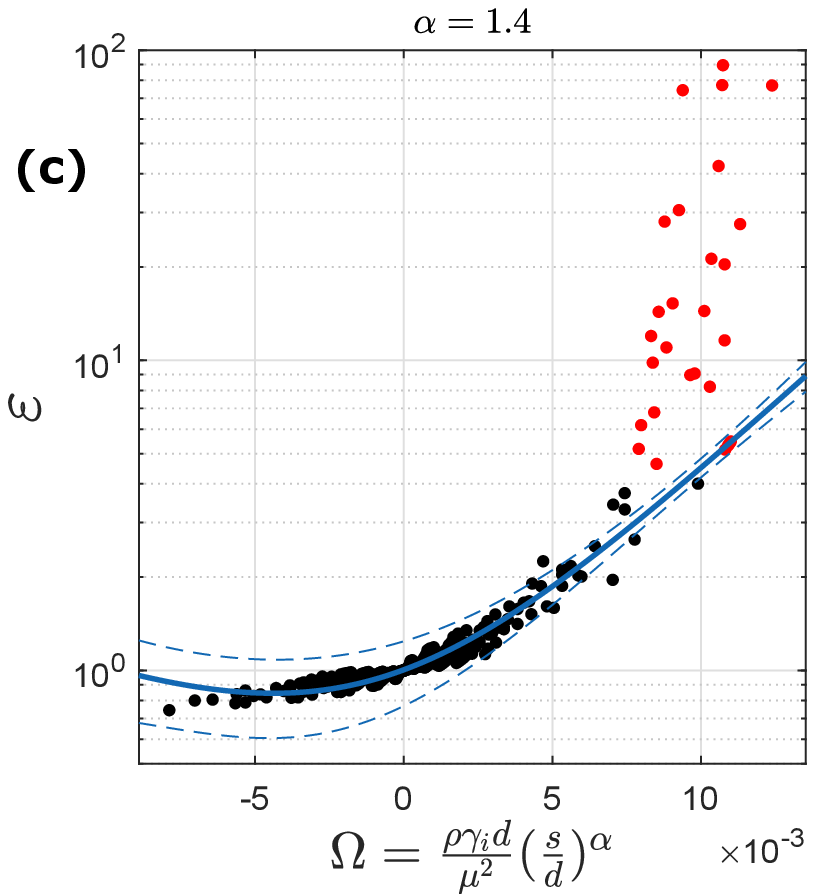}}
	\subfigure{\includegraphics[width=0.49\linewidth]{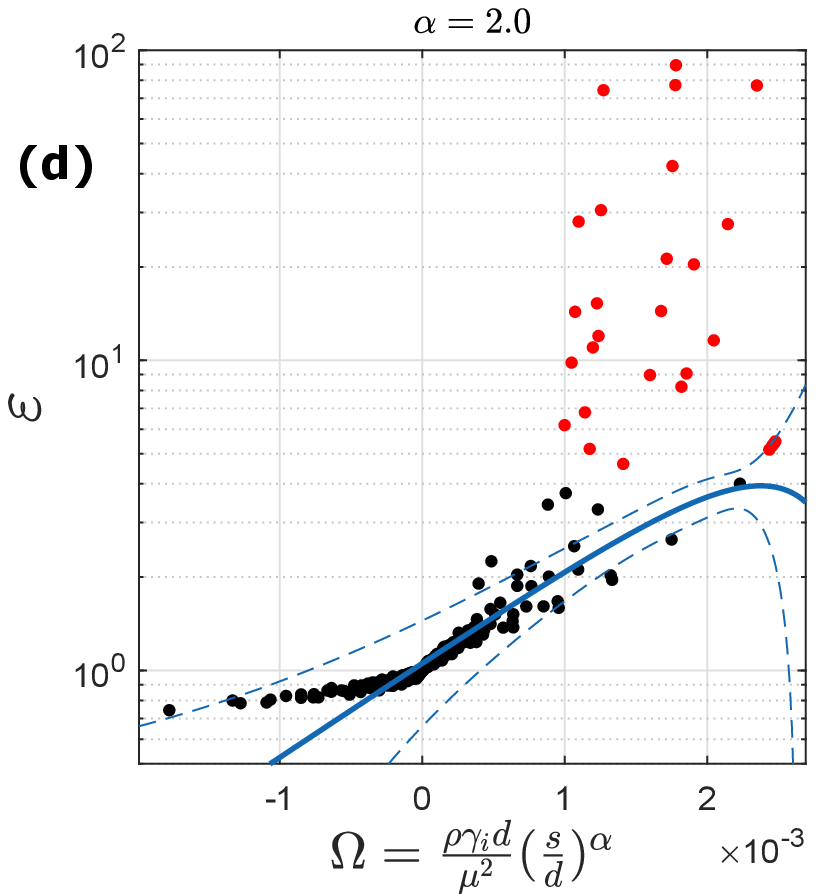}}
	\caption {Biexponential function fitted to the sub-critical portion of the $\varepsilon-\Omega$ values with $\alpha=1.0$ (a), 1.2 (b), 1.4 (c) and 2.0 (d) used in the definition of $\Omega=\frac{\rho \gamma_i d}{\mu^2}(\frac{s}{d})^\alpha$. The red data points are those in the critical regime which are excluded from the fitting. The fitted function appears in blue with the dashed lines indicating the prediction bounds with $95\%$ confidence.\label{fig:difAlphaFit}}
\end{figure}

\begin{figure}[!htbp]
	\centering
	\includegraphics[width=0.75\linewidth]{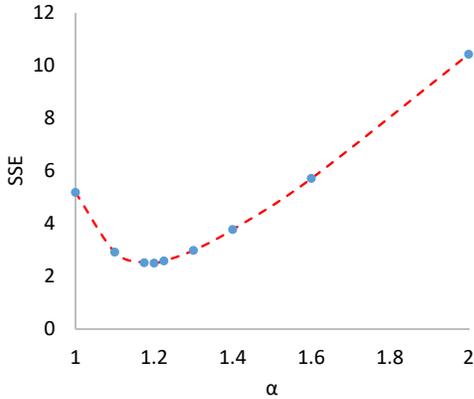}
	\caption {Sum of square differences/errors (SSE) versus $\alpha$ in $\Omega=\frac{\rho \gamma_i d}{\mu^2}(\frac{s}{d})^{\alpha}$. The value of $\alpha=1.2$ gives the minimum  SSE (best collapses all curves into a unified model). \label{fig:alpha-sse}}
\end{figure}

The data points can be fitted to a biexponential function up to the limit of the critical value of \textit{slip flow number}, $\Omega_{cr}$ beyond which the behaviour of $\varepsilon$ begins to deviate from any normal exponential function with a fast surge in the increase rate (see Fig. \ref{fig:difAlphaFit}). We select $\alpha$ in such a way to make all data (curves) best collapse into a single curve. The unified model is achieved by minimizing the sum of square errors (differences), SSE, for all datasets (Fig. \ref{fig:alpha-sse}). The value of $\alpha=1.2$ is found to produce the minimum SSE for the sub-critical portion of the results, giving the best convergence of all curves to a unified behaviour. The resulting $\Omega$ also provides a unified trend in critical regime where $\Omega$ exceeds the critical value, which suggests the universality of $\Omega$ with $\alpha$=1.2 in describing the nanofluidic slip flow regimes.    

\subsection{Interfacial energy values for nanotube materials}
Experimental reports on wetting properties of nanotube materials are reviewed in Table \ref{tab:gammai}.

\begin{table*}[]
	\centering
	\caption{\textbf{Wetting properties of nanotube materials reported in the literature.} Values of the interfacial energies are either reported directly in references (values with no parentheses) or calculated using Young’s equation based on the contact angle and the surface energy provided by each reference (values in parentheses). }
	\label{tab:gammai}
	\resizebox{\textwidth}{!}{%
		\begin{tabular}{|c|c|c|c|c|c|c|c|}
			\hline
			\textbf{Substance} & \textbf{layers/Walls} & \textbf{Method} & \textbf{\begin{tabular}[c]{@{}c@{}}Surface energy,\\ $\gamma_s$ ($mJ/m^2$)\end{tabular} } & \textbf{\begin{tabular}[c]{@{}c@{}}Substance-water\\ interfacial energy, \\$\gamma_i$ ($mJ/m^2$)\end{tabular}} & \textbf{\begin{tabular}[c]{@{}c@{}}Contact angle\\ of water,\\ $\theta$ $(degree)$\end{tabular}} & \textbf{Remarks} & \textbf{Ref.} \\ \hline
			Graphene    &      SL     &     Experiemntal      &    $115\pm4$       &     ($91.5\pm3$)      &      $71\pm1$     &       Chemical vapor deposition (CVD)   &      \cite{van2017direct}     \\ \hline
			Graphene    &      ML     &     Experiemntal      &    $119\pm3$       &     $83\pm7$      &      $70\pm1$     &   Direct measurement of $\gamma_i$, CVD    &      \cite{van2017direct}     \\ \hline
			Graphene    &      SL-ML     &     Experiemntal      &    $46.7$       &     ($90.5$)      &      $127\pm4$     &       -    &      \cite{wang2009wettability}     \\ \hline
			Graphene    &      ML     &     Experiemntal      &    -       &     -      &      $90.6-94.2$     &       -    &      \cite{rafiee2012wetting}          \\ \hline
			Graphene    &      SL-ML     &     Experiemntal      &    -       &     -      &      $93$     &      
			\begin{tabular}[c]{@{}c@{}}Independent of  number of layers\\ and substrate \\(Copper, SiO$_2$, Glass, HOPG)\end{tabular}
			
			&      \cite{raj2013wettability}     \\ \hline
			Graphene    &      -     &     \begin{tabular}[c]{@{}c@{}}Molecular\\dynamics\end{tabular}      &    -       & 
			\begin{tabular}[c]{@{}c@{}}$99.1^a$,\\$93.8^b$\end{tabular}   
			&      -     &
			\begin{tabular}[c]{@{}c@{}}$^a$ Ewald sum method,\\$^b$ Reaction field method\end{tabular} 
			&      \cite{dreher2019calculation}     \\ \hline
			Graphene & SL-ML & Experimental & - & - & 92 & Graphene on SiC substrate & \cite{shin2010surface} \\ \hline
			CNT & MW & Experimental & 45.3 & - & - & - & \cite{nuriel2005direct} \\ \hline
			CNT & Pristine MW & Experimental & 42.2 &  &  & - & \cite{roh2014characterization} \\ \hline
			Individual BNNT &    - &    Experimental &    26.7 &    21.1-25.8 &    85.4 $\pm$4.9 &    Wilhelmy method &    \cite{yum2006measurement}\\ \hline
			BN powder &    - &    Experimental &    43.8$^a$,
			66$^b$
			&    (19.2$^a$, 5.6$^b$) & 70$^a$, 33$^b$    &    $^a$, $^b$ Samples a and b &    \cite{rathod2004effect}\\ \hline
			Polycarbonate &    - &    Experimental &    38
			&    (38) & 90   &    $175$-$\mu m$-thick films &    \cite{bhurke2007surface}\\ \hline
			Polycarbonate &    - &    Experimental &    42.5$^a$,
			66$^b$
			&    (16.7$^a$, -0.6$^b$) & 69$^a$, 22.3$^b$    &    $^a$, $^b$ 1 and 120 min UV treatment &    \cite{ponter1994surface}\\ \hline
			Polysulfone
			&    - &    Experimental &    $45.57\pm4.45$
			&    ($48.93\pm4.5$) &92.65    &
			Rame-Hart contact angle goniometer
			&    \cite{ammar2017nanoclay}\\ \hline
			Polysulfone
			&    - &    Experimental &    39$^a$,
			65$^b$
			&    (10.9$^a$, 4.6$^b$) & 67$^a$, 33$^b$    &
			\begin{tabular}[c]{@{}c@{}}Ultrafiltration (UF) membrane\\$^a$, $^b$ 1 and 120 sec oxygen plasma \\treatment\end{tabular}
			&    \cite{kim2002surface}\\ \hline
			Polysulfone
			&    - &    Experimental &    42.56
			&    (22.71) &74    &
			-
			&    \cite{song2000surface}\\ \hline
		\end{tabular}%
	}
\end{table*}

\subsection{Bi-exponential and superexponential coefficients}
\begin{figure}[h!]
	\centering
	\includegraphics[width=0.75\linewidth]{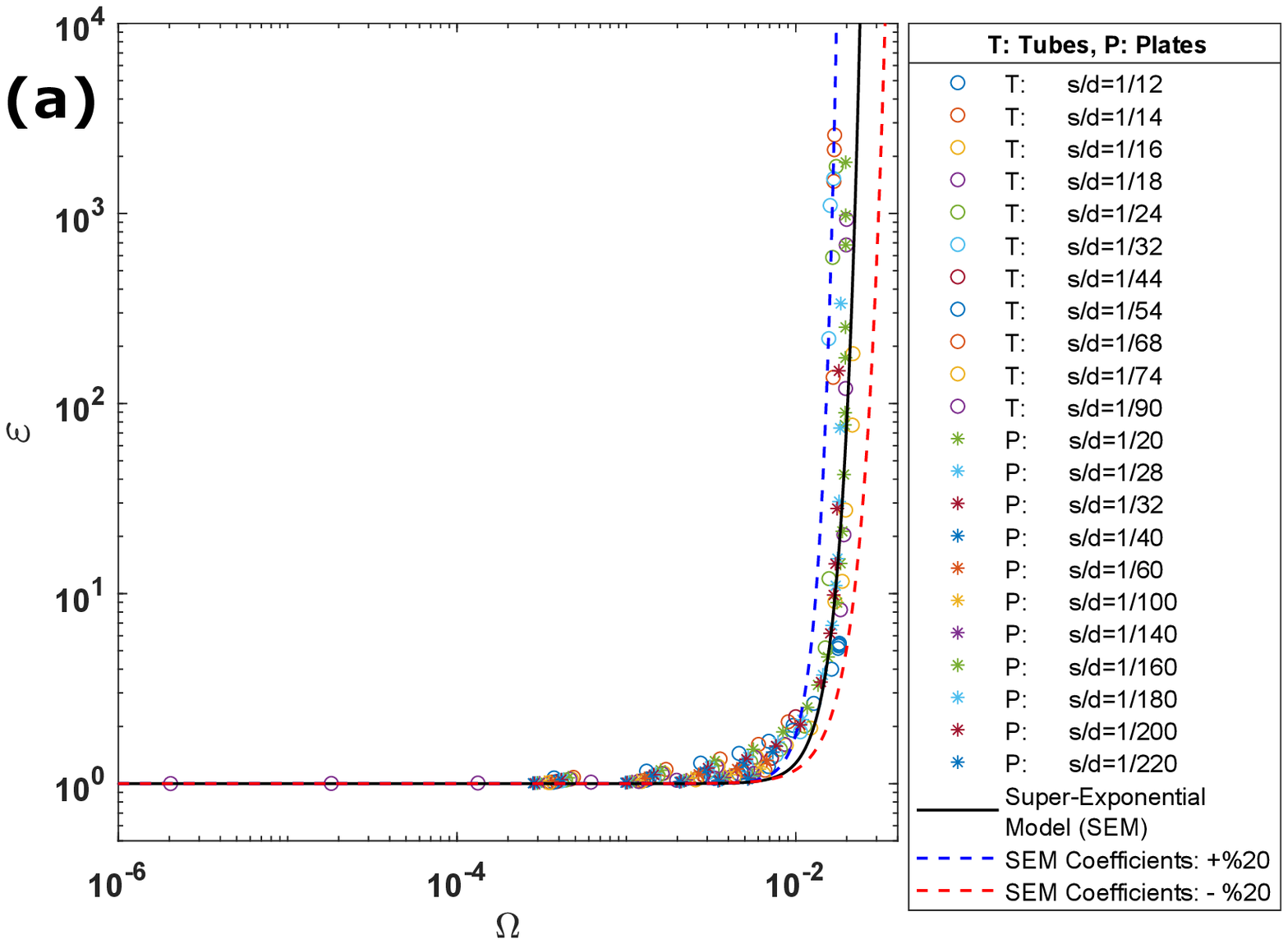}
	\includegraphics[width=0.75\linewidth]{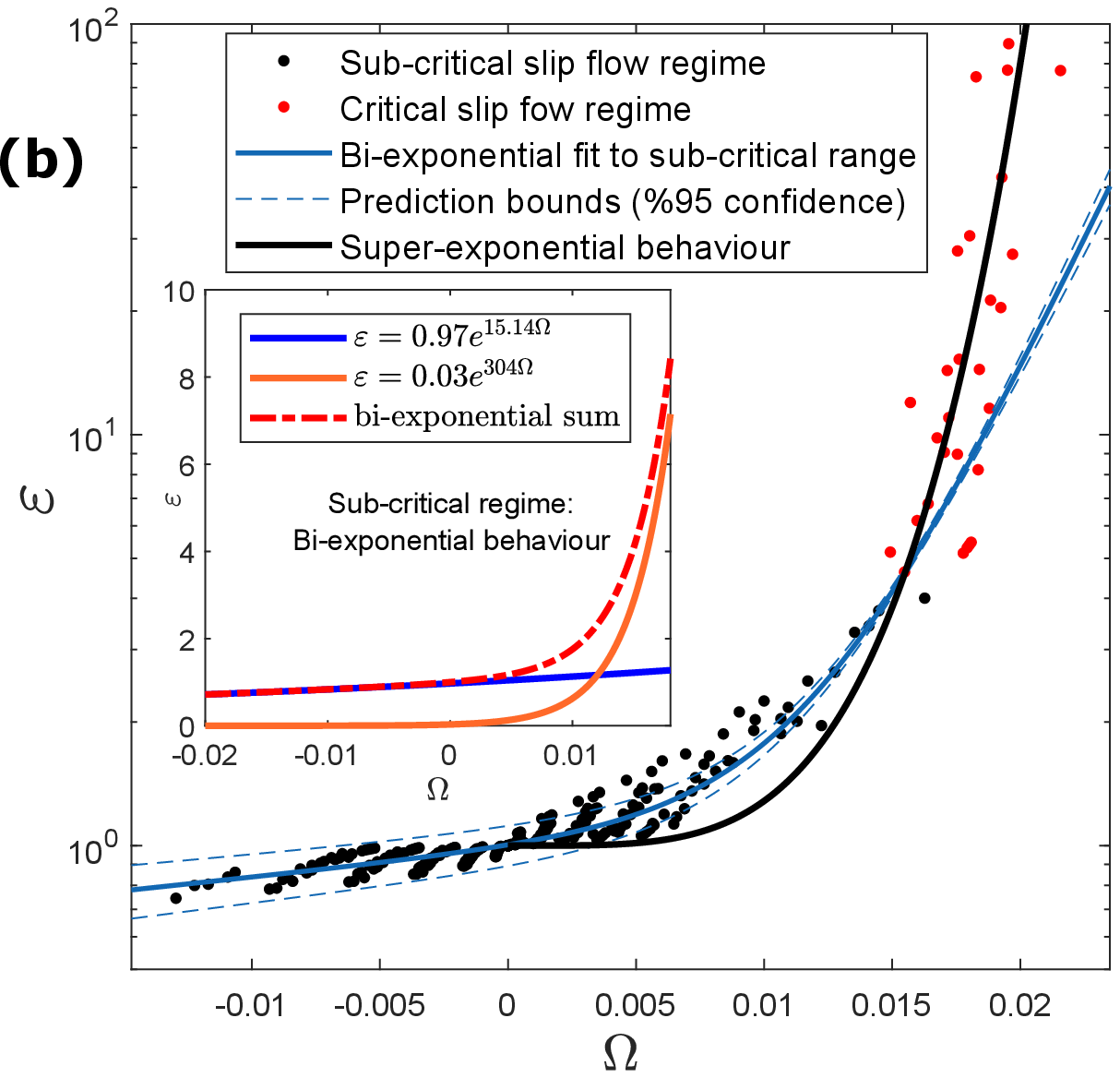}
	\caption {(a) Variation of the superexponential model coefficients. The optimized coefficients, \ie ($\varepsilon_0,\Omega_{cr},\beta$) are multiplied by $(1.2,0.8,1.2)$ and $(0.8,1.2,0.8)$ to determine a $+\%20$ and $-\%20$ variation of coefficients, respectively. (b) The superexponential slip flow model shown together with the bi-exponential behaviour of slip flow in the sub-critical regime. While the sub-critical behaviour is best characterized by the bi-exponential fit, with $\Omega$ exceeding the critical value ($\Omega_{cr}\approx0.0154$), a superexponential fit would be the best representative of the ultrafast slip flow behaviour. The inset illustrates the sum of exponential terms in the bi-exponential model for the sub-critical regime where the first term ($\varepsilon=0.97 e^{15.14 \Omega}$) best represents the flow attenuation behaviour (hydrophilic effects). The bi-exponential sum of both terms ($\varepsilon=0.97 e^{15.14 \Omega}+0.03 e^{304 \Omega}$) can properly fit the moderate flow enhancement behaviour (due to hydrophobic effects) in the sub-critical regime. \label{fig:super-bi fit}}
\end{figure}

In Fig. \ref{fig:super-bi fit}, we provide more details on the fitting coefficients of the sub-critical and critical slip flow regimes.

\subsection{Universality of the slip flow model ($\Omega-\varepsilon$ behaviour)}
Our model is proposed as a function of the nondimensional \textit{slip flow number} $\Omega$. To demonstrate the universality of this function, we conducted a large number of simulations for flow in tubes and between parallel plates with variable diameters and separation spaces. We also simulated a variety of cases, from repulsive to attractive solid-fluid forces, with different levels of interfacial energies. It was shown in Fig. 2 of the main paper that all cases can be unified into a single behaviour with our dimensionless \textit{slip flow number}, $\Omega$.

\begin{figure}[!htbp]
	\centering
	\includegraphics[width=\linewidth]{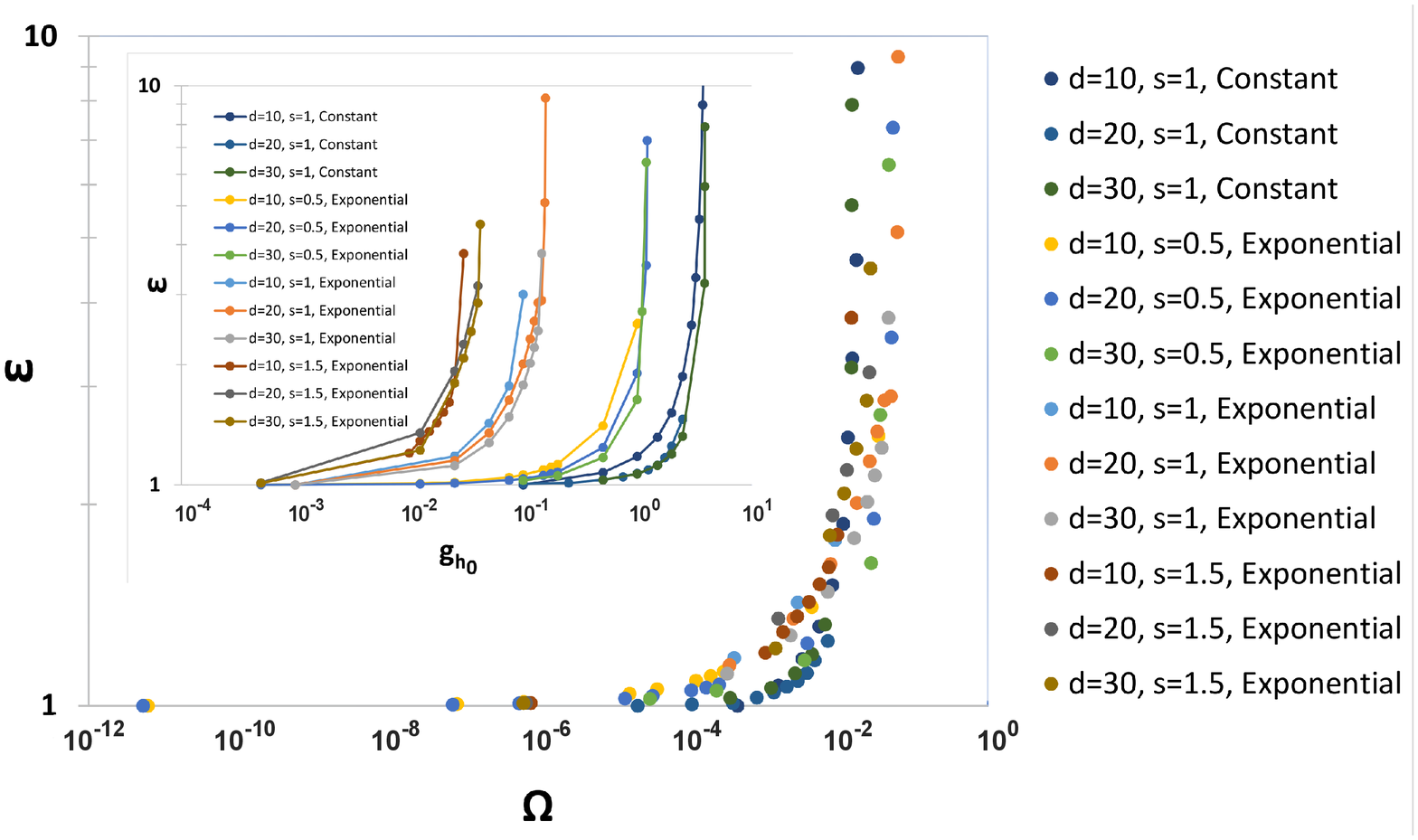}
	\caption {Universality of the proposed  slip flow model as a function of the \textit{slip flow number} $\Omega$. Here $\varepsilon$ versus $\Omega$ is shown for different force models used (constant or exponentially decaying with various decay lengths, $s$), all collapsing into a unified behaviour as given by the superexponential model. The proposed \textit{slip flow number} $\Omega$ is demonstrated to be a physically meaningful number in unifying the slip flow regime independent of the force models and their lengths of influence. The inset illustrates the same simulation data but shown as $\varepsilon$ versus $g_{h_0}$ -in lattice units- (where the constant force model is $\boldsymbol{F_h}(x)=-\rho(x)\boldsymbol{{g}_{h_0}}$, $0\!<\!x\!<\!s$ and exponential model is $\boldsymbol{F_h}(x)=-\rho(x)\boldsymbol{{g}_{h_0}}e^{-x/s}$
		with $x$ being the distance from walls). The inset shows the broad scatteredness of the slip flow behaviours in contrast with the single behaviour obtained using the \textit{slip flow number}. }\label{fig:universality}
\end{figure}

To further evaluate the universality of our model, we also implemented different force models (constant and exponentially-decaying: see Fig. \ref{fig:force_functions}). In addition, we change the decay length of the exponentially-decaying forces ($s$). The proposed \textit{slip flow number} $\Omega$ is shown to universally characterize the slip flow regimes with a unique behaviour for all cases under a wide range of conditions, as illustrated in Fig. \ref{fig:universality}.

\section{Discussions on the variation of model parameters in nanotube-water systems}
\subsection{Variation of viscosity, $\mu$ \label{app:mu}}
Indicated by many experiments \cite{goertz2007hydrophilicity,raviv2001fluidity} and MD simulations \cite{thomas2008reassessing,neek2016commensurability}, the viscosity of confined water varies spatially within an interface region near the walls where it is significantly affected by interfacial interaction. The interfacial region is determined as the water layer with a critical thickness next to the wall surface. The viscosity retains the value of bulk fluid for interior water beyond the critical thickness.  The critical thickness of the interfacial region is investigated by several studies (reviewed in Ref. \cite{wu2017wettability}), with a thickness of $0.7$ nm suggested as an average \cite{thomas2008reassessing,wu2017wettability}. 

Some researchers have suggested that the viscosity of water in the interface region could be a function of the contact angle \cite{wu2017wettability}. Fitted to some datasets from mainly MD simulations, the following linear relationship is proposed for the ratio of the interfacial viscosity to the viscosity for the bulk fluid, $\mu_i/\mu_\infty$ and the contact angle, $\theta$ ($degree$) \cite{wu2017wettability}:
\begin{equation}
\frac{\mu_i}{\mu_\infty}=-0.018 \theta+3.25.
\end{equation}
We argue that the use of a sole contact angle criterion based on MD simulations is inadequate for determining the interfacial characteristics of the nanofluidic systems. The interfacial energy states in addition to the contact angle -which is a tunable factor in MD simulations- could be a critical point to investigate the interface fluid properties in nanoscales \cite{leroy2015parametrizing}. In our model, the dynamic viscosity for the bulk fluid is used to calculate the \textit{slip flow number}, $\Omega$ while the sub-interface variations are intrinsically accounted for by the value of the interfacial energy. As viscosity can be a non-local property of the highly-confined fluid, care has been taken to ensure the transport coefficients are valid when selecting the simulation scales.

\subsection{Variation of interfacial layer characteristic thickness, $s$}
Although a conclusive law for hydrophobic interactions (between two hydrophobic surfaces in water) is yet to be achieved \cite{hammer2010search}, an exponentially decaying behaviour is usually observed in direct measurements \cite{israelachvili1982hydrophobic,tabor2013measurement,stock2015direct}. There has been also a long-lasting debate on the range of hydrophobic interactions. However, direct measurements of hydrophobic forces between two surfaces in aqueous solutions using the atomic force microscopy have revealed a pure hydrophobic interaction exponentially decaying over distances up to $10$-$20$ nm with a decay length of $0.3-2nm$, averagely $\sim1$nm \cite{meyer2006recent,hammer2010search,donaldson2014developing}. For the measurement of such pure hydrophobic forces, the surfaces are required to be smooth, continuous, free from defects and stable in water. 
The long-range attractive forces measured in larger ranges from $>200$ nm up to thousands of nm are also believed to be associated with surface preparation techniques  \cite{meyer2006recent}, and not the pure hydrophobicity. 
Overall, a typical value of $\sim 1$ nm is suggested as the decay length of both hydrophobic and hydrophilic interactions \cite{donaldson2014developing}, \ie the forces between two surfaces in water.  

The characteristic length of $1nm$ concluded for both hydrophobic and hydrophilic (solid-solid) interactions between two surfaces in water \cite{donaldson2014developing} could be assumed as an upper limit for $s$ which is the characteristic length of the solid-fluid effects (for a single surface in water). Those solid-fluid effects (which result in interfacial energies) are the origin of the solid-solid hydrophobic and hydrophilic interactions. In this study, considering the overlapping of hydrophobic effects, we assume a value of $0.5nm$ as $s$ (force characteristic length for a single solid surface in water ), equivalent to half of the decay length of hydrophobic and hydrophilic interactions (for two surfaces in water).  

\begin{figure}[h]
	\centering
	\includegraphics[width=\linewidth]{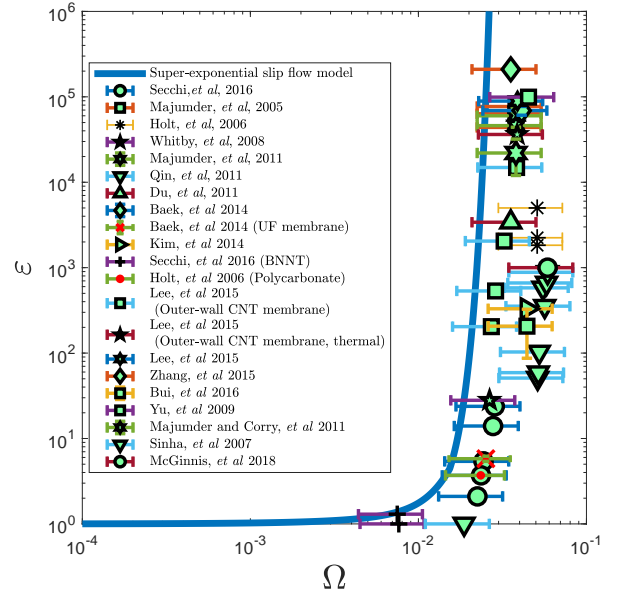}
	
	\caption {Superexponential model versus experimental data (as summarized in Table 	\ref{tab:exp-lit-data}) where data points correspond to the assumption of $s=0.75nm$ and errorbars correspond to the variation of $s$ in range of $0.5-1nm$.  
	}
	\label{fig:eps-om-validation-s}
\end{figure}

We also examine the validity of the model when considering the variations of the value of $s$, \ie the interfacial layer characteristic thickness (the solid-fluid force characteristic length). For this purpose, we plot our model versus the experimental values obtained from literature (as summarized in Table \ref{tab:exp-lit-data}), with a choice of $s$ in the range of $0.5-1nm$ (see Fig. \ref{fig:eps-om-validation-s}). As evident from the figure, the model predictions lie close to the data (varied with the assumed range of $s$), where $\Omega_{cr}$ obtained from simulations is within the same order suggested by data variations.

\subsection{Variation of density, $\rho$}
The variation of density is dominant in sub-$1.6$ nm dimensions where the sub-continuum transport is evident \cite{qin2011measurement,alexiadis2008self,alexiadis2008density}, extended up to $2.4$ nm diameters with the effects of first molecular layers \cite{alexiadis2008molecular,alexiadis2008density}. Nevertheless, while the density can increase or decrease up to a few folds in the sub-continuum regime or in the interfacial layers, the effective density in larger diameters is different from the bulk only in few percentages. The effective density can lead to the bulk value in upper-$2.4$ nm diameter tubes \cite{alexiadis2008molecular,alexiadis2008density}. Overall, the density variations are in much lower ranges than those of viscosity. Similar to viscosity, a bulk value is used in calculation of $\Omega$ in our model and density variations in interfacial layers are intrinsically accounted for in calculating the interfacial energies.


\section{Model Validation\label{App:C}}
\subsection{Slip flow numerical model vs analytical solutions} 
\begin{figure}[!htbp]
	\centering
	{\includegraphics[width=\linewidth]{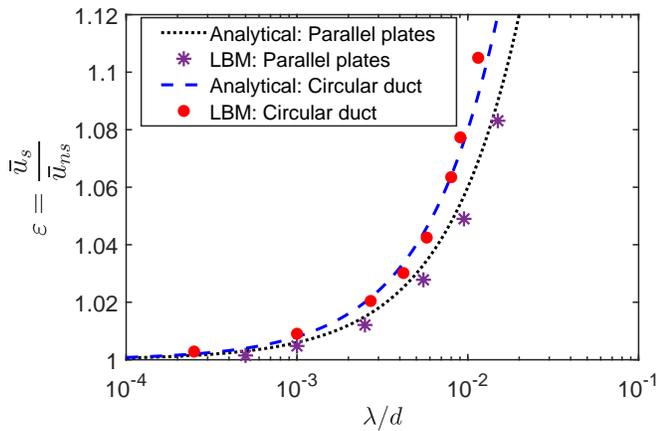}}
	\caption {Comparison of numerical results with analytical solutions for flow enhancement ratios versus normalized slip lengths. The analytical solutions are $\varepsilon=(1+6\lambda/d)$ for parallel plates (with separation distance of d) and $\varepsilon=(1+8\lambda/d)$ for circular ducts (with diameter d) \cite{halperin2005progress}.}
	\label{fig:Slipvalid}
\end{figure}

The validity of the numerical method in capturing the slip flow characteristics as induced by the fluid-solid wettability has been evaluated in previous studies \cite{aminpour2018slip,chen2014critical,harting2006lattice,benzi2006mesoscopic,zhu2005simulation,harting2010lattice} and also here against the analytical solution (Fig. \ref{fig:Slipvalid}). Under the influence of hydrophobic solid-fluid effects (forces resulting in interfacial energies), the velocity profiles deviate from those with a no-slip boundary wall assumption. Deviated velocity profiles result in increased mean velocity and an apparent slip length (when extrapolating the velocity profiles). Here, we implement the solid-fluid repulsive (hydrophobic) forces. Then the mean flow velocity and the apparent slip lengths (from extrapolated profiles) are calculated. In Fig. \ref{fig:Slipvalid}, we show that the velocity enhancements and corresponding slip lengths due to deviated velocity profiles are in agreement with the analytical relations available for the pipe flow and the flow between plates.

\newpage


\subsection{Numerical (MD) literature data for nanotube-water flow systems}    

A summary of the parameters used to produce Fig. \ref{fig:eps-d-validation}, extracted from Molecular Dynamics simulations on nanotubes reported in the literature is given in Table \ref{tab:lit-data-MD}.

\begin{table*}[h]
	\small
	\centering
	\caption{\textbf{MD simulation data from the literature on water flow in nanotubes}}
	\label{tab:lit-data-MD}
		\begin{tabular}{|c|c|c|c|c|c|c|}
			\hline
			\textbf{No.} & \textbf{Ref.} & \textbf{Method} & \textbf{Material} & \textbf{d (nm)} & \textbf{$\varepsilon$} & \textbf{Slip length (nm)} \\ \hline
			1 & Ritos, et al, 2014 \cite{ritos2014flow} & MD & CNT & 2.03 & 30.00 & 7.37 \\ \hline
			2 & Ritos, et al, 2014 \cite{ritos2014flow} & MD & BNNT & 2.07 & 8.50 & 1.94 \\ \hline
			3 & Ritos, et al, 2014 \cite{ritos2014flow} & MD & SiCNT & 2.06 & 5.50 & 1.16 \\ \hline
			4 & Thomas, et al, 2008 \cite{thomas2008reassessing} & MD & CNT & 1.66 & 432.47 & 89.53 \\ \hline
			5 & Thomas, et al, 2008 \cite{thomas2008reassessing} & MD & CNT & 2.22 & 198.70 & 54.86 \\ \hline
			6 & Thomas, et al, 2008 \cite{thomas2008reassessing} & MD & CNT & 2.77 & 110.39 & 37.88 \\ \hline
			7 & Thomas, et al, 2008 \cite{thomas2008reassessing} & MD & CNT & 3.33 & 81.82 & 33.64 \\ \hline
			8 & Thomas, et al, 2008 \cite{thomas2008reassessing} & MD & CNT & 3.88 & 53.25 & 25.34 \\ \hline
			9 & Thomas, et al, 2008 \cite{thomas2008reassessing} & MD & CNT & 4.44 & 53.25 & 29.00 \\ \hline
			10 & Thomas, et al, 2008 \cite{thomas2008reassessing} & MD & CNT & 4.99 & 48.05 & 29.35 \\ \hline
			11 & Joseph and Aluru, 2008 \cite{joseph2008carbon} & MD & CNT & 1.60 & 2052.00 & 410.20 \\ \hline
			12 & Joseph and Aluru, 2008 \cite{joseph2008carbon} & MD & BNNT & 1.60 & 1142.00 & 228.20 \\ \hline
			13 & Joseph and Aluru, 2008 \cite{joseph2008carbon} & MD & Si NT & 1.60 & 155.00 & 30.80 \\ \hline
			14 & Joseph and Aluru, 2008 \cite{joseph2008carbon} & MD & Rogh NT & 1.80 & 4.70 & 0.83 \\ \hline
			15 & Falk, et al, 2010 \cite{falk2010molecular} & MD & CNT & 10.13 & 92.30 & 115.61 \\ \hline
			16 & Falk, et al, 2010 \cite{falk2010molecular} & MD & CNT & 6.75 & 133.14 & 111.49 \\ \hline
			17 & Falk, et al, 2010 \cite{falk2010molecular} & MD & CNT & 5.40 & 199.01 & 133.65 \\ \hline
			18 & Falk, et al, 2010 \cite{falk2010molecular} & MD & CNT & 4.05 & 306.22 & 154.52 \\ \hline
			19 & Falk, et al, 2010 \cite{falk2010molecular} & MD & CNT & 2.68 & 534.24 & 178.64 \\ \hline
			20 & Falk, et al, 2010 \cite{falk2010molecular} & MD & CNT & 2.02 & 981.48 & 247.57 \\ \hline
			21 & Falk, et al, 2010 \cite{falk2010molecular} & MD & CNT & 1.60 & 1845.59 & 368.92 \\ \hline
			22 & Falk, et al, 2010 \cite{falk2010molecular} & MD & CNT & 1.34 & 2944.73 & 493.08 \\ \hline
			23 & Falk, et al, 2010  \cite{falk2010molecular} & MD & CNT & 1.07 & 5298.85 & 708.59 \\ \hline
			24 & Falk, et al, 2010 \cite{falk2010molecular} & MD & CNT & 0.93 & 23318.28 & 2710.63 \\ \hline
			25 & Falk, et al, 2010 \cite{falk2010molecular} & MD & CNT & 0.80 & 266194.13 & 26619.31 \\ \hline
			26 & Thomas, et al, 2010 \cite{thomas2010pressure} & MD & CNT & 7.00 & 34.74 & 29.50 \\ \hline
			27 & Thomas, et al, 2010 \cite{thomas2010pressure} & MD & CNT & 4.99 & 56.03 & 34.32 \\ \hline
			28 & Thomas, et al, 2010 \cite{thomas2010pressure} & MD & CNT & 4.44 & 65.44 & 35.76 \\ \hline
			29 & Thomas, et al, 2010 \cite{thomas2010pressure} & MD & CNT & 3.88 & 75.52 & 36.14 \\ \hline
			30 & Thomas, et al, 2010 \cite{thomas2010pressure} & MD & CNT & 3.33 & 90.33 & 37.18 \\ \hline
			31 & Thomas, et al, 2010 \cite{thomas2010pressure} & MD & CNT & 2.77 & 114.72 & 39.37 \\ \hline
			32 & Thomas, et al, 2010 \cite{thomas2010pressure} & MD & CNT & 2.22 & 191.87 & 52.97 \\ \hline
			33 & Thomas, et al, 2010 \cite{thomas2010pressure} & MD & CNT & 1.66 & 443.49 & 91.82 \\ \hline
			34 & Du, et al, 2011 \cite{du2011membranes} & MD & CNT & 4.00 & 520.00 & 259.50 \\ \hline
			35 & Babu, et al, 2011 \cite{babu2011role} & MD & CNT & 5.42 & 1.49 & 0.33 \\ \hline
			36 & Babu, et al, 2011 \cite{babu2011role} & MD & CNT & 5.16 & 1.72 & 0.46 \\ \hline
			37 & Babu, et al, 2011 \cite{babu2011role} & MD & CNT & 4.88 & 2.00 & 0.61 \\ \hline
			38 & Babu, et al, 2011 \cite{babu2011role} & MD & CNT & 4.33 & 2.82 & 0.98 \\ \hline
			39 & Babu, et al, 2011 \cite{babu2011role} & MD & CNT & 4.07 & 3.37 & 1.20 \\ \hline
			40 & Babu, et al, 2011 \cite{babu2011role} & MD & CNT & 3.79 & 4.10 & 1.47 \\ \hline
			41 & Babu, et al, 2011 \cite{babu2011role} & MD & CNT & 3.52 & 5.05 & 1.79 \\ \hline
			42 & Babu, et al, 2011 \cite{babu2011role} & MD & CNT & 3.25 & 6.34 & 2.17 \\ \hline
			43 & Babu, et al, 2011 \cite{babu2011role} & MD & CNT & 2.98 & 8.17 & 2.67 \\ \hline
			44 & Babu, et al, 2011 \cite{babu2011role} & MD & CNT & 2.50 & 13.48 & 3.89 \\ \hline
			45 & Babu, et al, 2011 \cite{babu2011role} & MD & CNT & 2.16 & 20.21 & 5.19 \\ \hline
			46 & Babu, et al, 2011 \cite{babu2011role} & MD & CNT & 1.93 & 27.81 & 6.48 \\ \hline
			47 & Babu, et al, 2011 \cite{babu2011role} & MD & CNT & 1.35 & 64.24 & 10.68 \\ \hline
			48 & Nicholls, et al, 2012 \cite{nicholls2012water} & MD & CNT & 0.96 & 860.00 & 103.08 \\ \hline
			49 & Kannam, et al, 2013 \cite{kannam2013fast} & MD & CNT & 6.52 & 95.05 & 76.59 \\ \hline
			50 & Kannam, et al, 2013 \cite{kannam2013fast} & MD & CNT & 4.87 & 155.85 & 87.11 \\ \hline
		\end{tabular}%
\end{table*}

\begin{table*}[]
	\renewcommand\thetable{III}
	\centering
	\caption{\textbf{Continued from previous page: MD simulation data from the literature on water flow in nanotubes}}
	\label{tab:lit-data-MD}
		\begin{tabular}{|c|c|c|c|c|c|c|}
			\hline
			\textbf{No.} & \textbf{Ref.} & \textbf{Method} & \textbf{Material} & \textbf{d (nm)} & \textbf{$\varepsilon$} & \textbf{Slip length (nm)} \\ \hline    
			51 & Kannam, et al, 2013 \cite{kannam2013fast} & MD & CNT & 4.34 & 181.28 & 92.81 \\ \hline
			52 & Kannam, et al, 2013 \cite{kannam2013fast} & MD & CNT & 3.80 & 206.70 & 98.10 \\ \hline
			53 & Kannam, et al, 2013 \cite{kannam2013fast} & MD & CNT & 3.26 & 260.48 & 100.92 \\ \hline
			54 & Kannam, et al, 2013 \cite{kannam2013fast} & MD & CNT & 2.71 & 337.44 & 121.82 \\ \hline
			55 & Kannam, et al, 2013 \cite{kannam2013fast} & MD & CNT & 2.16 & 517.50 & 131.63 \\ \hline
			56 & Kannam, et al, 2013 \cite{kannam2013fast} & MD & CNT & 1.90 & 664.24 & 157.49 \\ \hline
			57 & Kannam, et al, 2013 \cite{kannam2013fast} & MD & CNT & 1.63 & 911.49 & 174.72 \\ \hline
			58 & Casanova, et al, 2019 \cite{casanova2018surface} & MD & Pristine CNT & 4.07 & 105.15 & 53.00 \\ \hline
			59 & Casanova, et al, 2019 \cite{casanova2018surface} & MD & Turbostratic CNT & 4.07 & 20.07 & 9.70 \\ \hline
			60 & Casanova, et al, 2019 \cite{casanova2018surface} & MD & carbon nitride nanotube & 4.07 & \textless{}2.97 & \textless{}1 \\ \hline
			61 & Bordin, et al, 2013 \cite{bordin2013relation} & MD & CNT & 1.93 & 17.71 & 4.03 \\ \hline
			62 & Bordin, et al, 2013 \cite{bordin2013relation} & MD & CNT & 1.79 & 17.78 & 3.75 \\ \hline
			63 & Bordin, et al, 2013 \cite{bordin2013relation} & MD & CNT & 1.65 & 20.71 & 4.06 \\ \hline
			64 & Bordin, et al, 2013 \cite{bordin2013relation} & MD & CNT & 1.45 & 20.10 & 3.47 \\ \hline
			65 & Bordin, et al, 2013 \cite{bordin2013relation} & MD & CNT & 1.37 & 23.72 & 3.88 \\ \hline
			66 & Bordin, et al, 2013 \cite{bordin2013relation} & MD & CNT & 1.24 & 27.36 & 4.08 \\ \hline
			67 & Bordin, et al, 2013 \cite{bordin2013relation} & MD & CNT & 1.17 & 23.82 & 3.34 \\ \hline
			68 & Bordin, et al, 2013 \cite{bordin2013relation} & MD & CNT & 1.00 & 49.62 & 6.07 \\ \hline
			69 & Bordin, et al, 2013 \cite{bordin2013relation} & MD & CNT & 0.90 & 35.39 & 3.87 \\ \hline
			70 & Bordin, et al, 2013 \cite{bordin2013relation} & MD & CNT & 0.82 & 53.29 & 5.36 \\ \hline
			71 & Bordin, et al, 2013 \cite{bordin2013relation} & MD & CNT & 0.68 & 62.65 & 5.22 \\ \hline
			72 & Bordin, et al, 2013 \cite{bordin2013relation} & MD & CNT & 0.51 & 102.02 & 6.48 \\ \hline
			73 & Bordin, et al, 2013 \cite{bordin2013relation} & MD & CNT & 0.48 & 118.46 & 7.03 \\ \hline
			74 & Bordin, et al, 2013 \cite{bordin2013relation} & MD & CNT & 0.43 & 124.92 & 6.72 \\ \hline
			75 & Bordin, et al, 2013 \cite{bordin2013relation} & MD & CNT & 0.41 & 211.35 & 10.88 \\ \hline
			76 & Bordin, et al, 2013 \cite{bordin2013relation} & MD & CNT & 0.38 & 277.80 & 13.13 \\ \hline
			77 & Walther, et al, 2013 \cite{walther2013barriers} & MD & CNT & 2.00 & 253.00 & 63.00 \\ \hline
			78 & Tao, et al, 2018 \cite{tao2018confinement} & MD & CNT & 101.07 & 1.00 & 0.00 \\ \hline
			79 & Tao, et al, 2018 \cite{tao2018confinement} & MD & CNT & 4.71 & 230.00 & 134.76 \\ \hline
			80 & Tao, et al, 2018 \cite{tao2018confinement} & MD & CNT & 3.73 & 332.73 & 154.57 \\ \hline
						81 & Tao, et al, 2018 \cite{tao2018confinement} & MD & CNT & 2.89 & 481.54 & 173.57 \\ \hline
			82 & Tao, et al, 2018 \cite{tao2018confinement} & MD & CNT & 2.10 & 1046.83 & 274.76 \\ \hline
			83 & Tao, et al, 2018 \cite{tao2018confinement} & MD & CNT & 1.79 & 1285.87 & 287.89 \\ \hline
			84 & Tao, et al, 2018 \cite{tao2018confinement} & MD & CNT & 1.53 & 1783.96 & 340.71 \\ \hline
			85 & Tao, et al, 2018 \cite{tao2018confinement} & MD & CNT & 1.26 & 2912.73 & 459.67 \\ \hline
			86 & Tao, et al, 2018 \cite{tao2018confinement} & MD & CNT & 1.00 & 5161.59 & 645.07 \\ \hline
			87 & Tao, et al, 2018 \cite{tao2018confinement} & MD & CNT & 0.72 & 10776.59 & 969.38 \\ \hline
			88 & Sam, et al, 2019 \cite{sam2019water} & MD & CNT (zigzag) & 4.89 & 166.47 & 101.11 \\ \hline
			89 & Sam, et al, 2019 \cite{sam2019water} & MD & CNT (zigzag) & 4.34 & 185.60 & 100.15 \\ \hline
			90 & Sam, et al, 2019 \cite{sam2019water} & MD & CNT (zigzag) & 3.78 & 218.66 & 102.89 \\ \hline
			91 & Sam, et al, 2019 \cite{sam2019water} & MD & CNT (zigzag) & 3.23 & 251.73 & 101.34 \\ \hline
			92 & Sam, et al, 2019 \cite{sam2019water} & MD & CNT (zigzag) & 2.72 & 326.63 & 110.54 \\ \hline
			93 & Sam, et al, 2019 \cite{sam2019water} & MD & CNT (zigzag) & 2.43 & 362.67 & 109.92 \\ \hline
			94 & Sam, et al, 2019 \cite{sam2019water} & MD & CNT (zigzag) & 2.15 & 443.31 & 118.72 \\ \hline
			95 & Sam, et al, 2019 \cite{sam2019water} & MD & CNT (zigzag) & 1.88 & 537.90 & 126.39 \\ \hline
			96 & Sam, et al, 2019 \cite{sam2019water} & MD & CNT (zigzag) & 1.59 & 718.88 & 142.57 \\ \hline
			97 & Sam, et al, 2019 \cite{sam2019water} & MD & CNT (zigzag) & 1.36 & 933.35 & 157.96 \\ \hline
		\end{tabular}%
\end{table*}

\end{document}